%% file: main.tex
\documentclass[preprint]{elsarticle} 

\makeatletter
\def\ps@pprintTitle{%
\let\@oddhead\@empty
\let\@evenhead\@empty
\def\@oddfoot{\centerline{\thepage}}%
\let\@evenfoot\@oddfoot}
\makeatother

\usepackage{etoolbox}
\newcommand{\preprintdate}{July 14, 2026}
\patchcmd{\MaketitleBox}{\footnotesize\itshape\elsaddress\par\vskip36pt}{\footnotesize\itshape\elsaddress\par\parbox[b][36pt]{\linewidth}{\vfill\hfill\textnormal{\preprintdate}\hfill\null\vfill}}{}{}%
\patchcmd{\pprintMaketitle}{\footnotesize\itshape\elsaddress\par\vskip36pt}{\footnotesize\itshape\elsaddress\par\parbox[b][36pt]{\linewidth}{\vfill\hfill\textnormal{\preprintdate}\hfill\null\vfill}}{}{}%

\usepackage[
colorlinks=true,%
breaklinks=true,%
linkcolor=blue,
urlcolor=blue,%
citecolor=blue,%
pdftitle={Fine-Grained Open-Set Fault Diagnosis via Metric-Guided Time-Frequency Configuration Selection and Class-Specific Autoencoders}, 
pdfkeywords={Prognostics and health management, Deep learning, Open-set fault diagnosis, Class-specific autoencoder, Rotating machinery},	
pdfauthor={Youngjae Jeon; Dongjin Lee},		
bookmarksopen=false,
pdfpagemode=UseNone]{hyperref}
\usepackage{multicol}
\usepackage{algorithm}
\usepackage{algorithmicx}
\usepackage{algpseudocode}
\usepackage{tabularx}
\usepackage{array}
\usepackage{multirow}
\usepackage{threeparttable}
\usepackage{booktabs}
\usepackage{graphicx}
\usepackage{arydshln}
\usepackage{setspace}   
\usepackage{xcolor}
\usepackage[mathlines,running]{lineno}
\usepackage{epstopdf}
\usepackage{graphicx}

\input{ListPackages}

\newcommand{\RNum}[1]{\uppercase\expandafter{\romannumeral #1\relax}}
\usepackage{amsthm}
\makeatletter
\def\els@aparagraph[#1]#2{\elsparagraph[#1]{#2\@addpunct{.}}}
\def\els@bparagraph#1{\elsparagraph*{#1\@addpunct{.}}}
\makeatother
\input{mymacros.tex}

\graphicspath{{./figures/}}
\begin{document}	
\begin{frontmatter}

\title{Fine-Grained Open-Set Fault Diagnosis via Metric-Guided Time-Frequency Configuration Selection and Class-Specific Autoencoders}
		
		
\author[HYU1]{Youngjae Jeon}
\ead{yjeon@hanyang.ac.kr}
\author[HYU2]{Dongjin Lee\corref{cor1}}
\ead{dlee46@hanyang.ac.kr}

\cortext[cor1]{Corresponding author}
\address[HYU1]{Department of Automotive Engineering (Automotive-Computer Convergence), \\ 
Hanyang University, Seoul, South Korea}
\address[HYU2]{Department of Automotive Engineering, Hanyang University, Seoul, South Korea}

\begin{abstract} 

Reliable fault diagnosis of rotating machinery is essential for the safe and stable operation of industrial systems. Although deep learning methods perform well under closed-set conditions, real machinery may encounter previously unseen fault states. Existing open-set fault diagnosis (OSFD) methods remain limited in fine-grained severity diagnosis because they often rely on coarse type levels, heuristically selected Short-Time Fourier Transform (STFT) settings, and global class boundaries. We propose a fine-grained OSFD method that combines metric-guided data-centric (MGDC) STFT configuration selection with class-specific autoencoder (CSAE)-based anomaly rejection. MGDC screens candidate STFT configurations using the Silhouette score computed from spectrogram representations, identifying promising time-frequency representations before network training. The diagnostic model then uses a bank of CSAEs to learn compact class-specific manifolds for known degradation states. During inference, reconstruction-error-based class affinity identifies known classes, while a dual-criteria mechanism based on latent dimension-wise boundaries and class-specific reconstruction error rejects unknown samples. Experiments on the Case Western Reserve University (CWRU) and Paderborn University (PU) bearing datasets show that the proposed method achieves H-scores of $0.9924$ and $0.9509$ for fine-grained fault severity diagnosis.  MGDC also identifies the best-performing configuration found by exhaustive search while evaluating only $9$ of $38$ candidates on CWRU and $2$ of $39$ candidates on PU, reducing the selection cost by factors of $5.69$ and $29.87$, respectively. These results indicate that the proposed method supports accurate open-set severity diagnosis with substantially lower configuration-selection cost.

\end{abstract}


\end{frontmatter}


\section{Introduction} \label{sec:Intro}

Fault diagnosis of rotating machinery is essential for ensuring the reliability, safety, and continuous operation of modern industrial systems, including wind turbines, electric vehicles, aerospace propulsion units, and high-speed manufacturing lines~\citep{lei2020applications, park2025spectral, hu2024transferable, bao2026theory}. Data-driven methods, and deep learning in particular, have reshaped this diagnostic paradigm over the past decade~\citep{zhao2019deep, han2021hybrid}: by extracting complex, nonlinear fault signatures directly from raw condition-monitoring signals, such models classify mechanical health conditions with high accuracy~\citep{lei2020applications, guo2026fault, xu2025lightweight, bao2026theory, liu2025mitigating}.

A key limitation of these approaches, however, is their closed-set assumption---the premise that the training and testing data share an identical class distribution, so that test labels are restricted to the known fault classes seen during training. Real-world industrial machinery routinely encounters states absent from the offline training set: 
unpredictable operating conditions, unobserved fault types, or novel degradation levels. In such open-set settings, a closed-set network maps every input to one of its known classes by construction, and therefore silently misclassifies unknown faults with high confidence. Open-set fault diagnosis (OSFD) addresses this failure mode by preserving closed-set classification accuracy while explicitly rejecting unobserved anomalies~\citep{liu2025mitigating, rehman2023open, kim2026center,han2022out}.

OSFD relaxes the closed-set assumption by modeling an additional unknown-class region during inference, but its deployment exposes a second, subtler bottleneck: the distinction between coarse-grained fault type and fine-grained fault severity~\citep{lang2024coarse}. We use coarse-grained fault type to denote the macroscopic failure mode (inner race, outer race, ball), and fine-grained fault severity to denote progressive degradation within a single fault type (e.g., the 0.007/0.014/0.021 inch defect diameter classes of the CWRU benchmark~\citep{smith2015rolling}). Most existing OSFD studies are designed and evaluated for type separation alone--- distinguishing, say, an inner race fault from a rolling element. Because each macroscopic type carries distinct characteristic defect frequencies set by bearing kinematics, these types form well-separated clusters in the latent space, which makes the task comparatively forgiving. However, identifying the fault type is often insufficient in practice: operators must track a component's progressive degradation to estimate remaining useful life, schedule predictive maintenance, and prevent catastrophic failure. Transitioning from coarse-grained type diagnosis to fine-grained severity diagnosis is thus essential, and it places a markedly stricter demand on the diagnostic model's discriminative capacity.

This demand arises because adjacent severity stages of a single fault type share identical characteristic defect frequencies. They differ only marginally in localized spectral energy distributions, harmonic modulations, or the amplitude of time-varying transient impacts across the time-frequency domain. Since these non-stationary variations are highly microscopic, the inter-class variance is correspondingly narrow. Consequently, existing OSFD approaches tuned for type diagnosis degrade sharply when applied to severity diagnosis. Even representative open-set classifiers (e.g., OpenMax~\citep{bendale2016towards, ye2026adaptive}, CPL~\citep{yang2020convolutional}, ARPL~\citep{chen2021adversarial}) fail to place accurate decision boundaries when the initial signal processing stage does not capture and preserve these fine time-frequency details. The success of a fine-grained OSFD method therefore depends on the quality of the initial input representation.

Most existing OSFD pipelines select Short-Time Fourier Transform (STFT) hyperparameters by trial-and-error~\citep{orhan2025comparative, hwang2025frequency, fu2023magva}. The STFT remains among the most widely adopted techniques for processing non-stationary vibration signals~\citep{orhan2025comparative}. However, its fundamental hyperparameters, particularly the frequency bins and time steps, are frequently chosen empirically rather than by objective optimization~\citep{nisar2016efficient}. Since the STFT is bound by a strict time-frequency resolution trade-off, a suboptimal STFT configuration induces spectral leakage and energy dispersion that obscure the localized harmonic variations distinguishing adjacent severity stages. Consequently, the downstream diagnostic model receives an ambiguous time-frequency representation in which severity-discriminative spectral patterns are partially obscured, and fine-grained misclassifications follow. Since this heuristic preprocessing bottleneck undermines the reliability of the entire predictive-maintenance pipeline, an objective, data-driven method for constructing time-frequency representations is needed~\citep{ni2022fault,gui2026motor}.

To isolate known conditions from unknown anomalies, recent OSFD methods fall into four main streams: discriminative boundary-based, density- and generative-based, uncertainty- and evidential-based, and hybrid multi-constraint methods~\citep{rehman2023open}. These paradigms separate macroscopically distinct fault types effectively, but they share a critical structural vulnerability on the dense, continuous manifolds of fine-grained severity: the failure of multidimensional geometric constraints.
Whether they use distance metrics, reconstruction errors, or probabilistic uncertainties, these approaches compress directional fault signatures into a single scalar metric. For instance, discriminative approaches based on extreme value theory or metric learning~\citep{yu2021deep, ye2026adaptive, chopra2025open, he2021deep} typically enforce isotropic radial boundaries. Even when adopting the Mahalanobis distance~\citep{lee2018simple, denouden2018improving}, they inherently rely on a strict multivariate Gaussian assumption that deep networks frequently violate~\citep{venkataramanan2023gaussian, mueller2025mahalanobis}. Reconstructive models such as global autoencoders rely on a single global reconstruction error, which masks localized fault signatures and over-generalizes. Evidential approaches~\citep{wei2025open} encounter the same bottleneck, compressing a multidimensional Dirichlet output into a single uncertainty score. A scalar anomaly score discards the dimension-specific directional information needed to separate fine-grained severity states, which entangles the manifolds and weakens unknown rejection.
In fine-grained diagnosis, individual diagnostic paradigms often struggle to resolve the entangled manifolds. Combining complementary approaches into a unified method compensates for the limitations of each: pairing reconstruction-based representation learning with explicit multi-dimensional boundary modeling preserves directional fault information while supplying a more reliable rejection criterion, yielding tighter decision boundaries for fine-grained OSFD.

To address the heuristic bottleneck and structural blind spots, we propose an OSFD method tailored to fine-grained severity. This method integrates a metric-guided data-centric preprocessing strategy with a hybrid OSFD architecture.
First, the MGDC preprocessing strategy resolves the heuristic preprocessing bottleneck in STFT configuration determination. Instead of running computationally expensive hyperparameter optimization, MGDC scores candidate STFT configurations before network training with a clustering-based separability criterion, the Silhouette score. This criterion identifies an effective time-frequency representation at a fraction of the usual search cost.
Second, we develop a class-specific autoencoder (CSAE) architecture for fine-grained OSFD. This architecture comprises a shared feature extractor, a bank of CSAEs, a classifier, and a dual-criteria anomaly rejection mechanism. The shared feature extractor isolates non-stationary defect signatures from the time-frequency domain; the CSAE bank then processes these representations, with each autoencoder securing a tight discriminative margin by penalizing cross-class reconstructions while minimizing reconstruction error on its target class. The reconstruction error-based classifier assigns a known fault state from the resulting per-class errors. 
The rejection mechanism then uses both the latent representation and the reconstruction error to detect unknown fault states. During training, the latent representations of CSAEs define a class-specific manifold and a multi-dimensional boundary. This boundary verifies the latent representations across specific spatial axes, avoiding the dimensional compression and over-generalization of a global scalar metric. Operating under the fundamental premise that an autoencoder (AE) yields magnified errors on out-of-distribution samples, a strict reconstruction error limit imposes another rejection criterion. By coupling high-fidelity input representations with complementary diagnostic criteria, the proposed method establishes reliable decision boundaries for rejecting unknown fault states in fine-grained OSFD.
The main contributions of this study are summarized as follows:
\begin{itemize}
    \item We introduce an MGDC preprocessing strategy that selects an effective STFT configuration. It uses the Silhouette score to rank candidate STFT configurations before network training. This replaces the conventional heuristic search with an objective criterion and avoids exhaustive training over the full search space.
    \item We develop a hybrid diagnostic architecture built on a bank of CSAEs to reduce feature entanglement among fine-grained severity states. Dedicating each autoencoder to a specific known severity class promotes compact class-wise representations and improves separability among adjacent degradation states.
    \item We propose a robust dual-criteria anomaly rejection mechanism that combines a multi-dimensional latent representation boundary and a reconstruction error constraint. This mechanism overcomes the dimensional compression  of global scalar metrics and yields a more reliable decision boundary for rejecting unobserved anomalies.
\end{itemize}

The remainder of this paper is organized as follows: Section~\ref{sec:Theoretical background} presents the theoretical background. Section~\ref{sec:Method} introduces the proposed CSAE method for fine-grained OSFD. The results are discussed in Section~\ref{sec:Results}. Finally, the conclusion is provided in Section~\ref{sec:Conclusion}.


\section{Theoretical background} \label{sec:Theoretical background} 
This section presents the theoretical background of the proposed method. Section~\ref{sec:Fine-grained} formulates fine-grained severity-based OSFD and discusses the limitations of closed-set classifiers, global autoencoders, and coarse fault-type labels. Section~\ref{sec:Silhouette} introduces the Silhouette score as a data-centric measure for evaluating the separability of candidate time-frequency representations. These discussions define the three problems addressed by the proposed method.

\subsection{Fine-grained severity-based open-set fault diagnosis} \label{sec:Fine-grained}

\begin{figure}[ht!]
\centering
\begin{subfigure}[b]{0.48\textwidth} 
    \includegraphics[width=\textwidth]{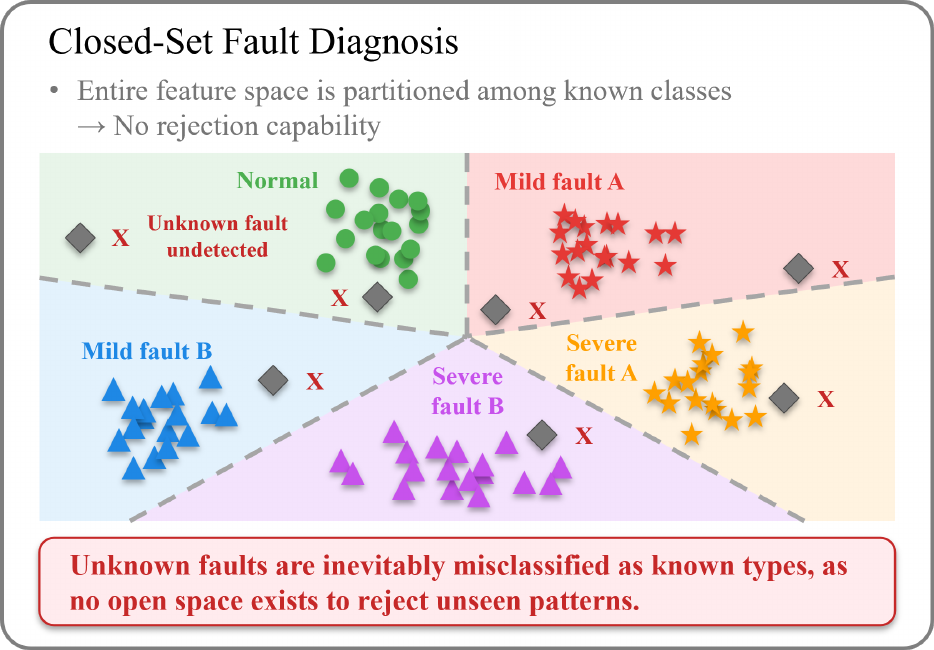}
    \caption{}
    \label{fig:CSFD}
\end{subfigure}
\begin{subfigure}[b]{0.48\textwidth} 
    \includegraphics[width=\textwidth]{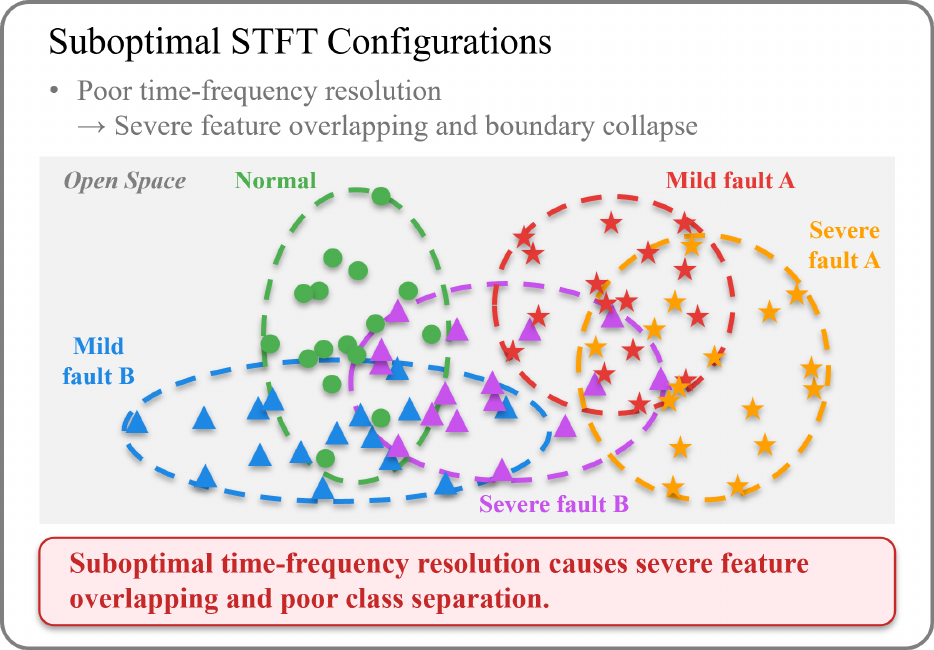}
    \caption{} 
    \label{fig:Suboptimal} 
\end{subfigure}
\par\bigskip
\begin{subfigure}[b]{0.48\textwidth} 
    \includegraphics[width=\textwidth]{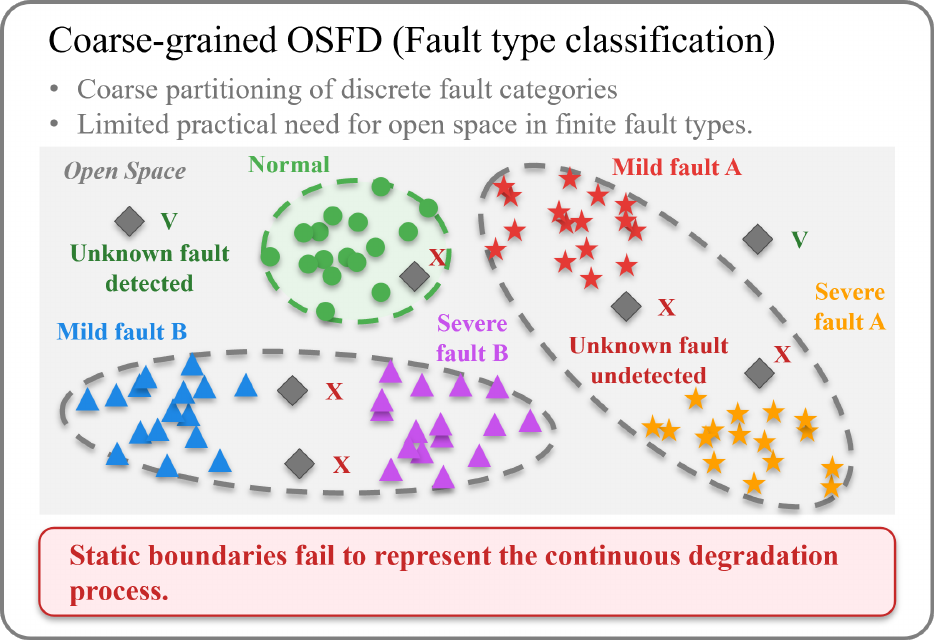}
    \caption{} \label{fig:Coarse_OSFD} 
\end{subfigure}
\begin{subfigure}[b]{0.48\textwidth} 
    \includegraphics[width=\textwidth]{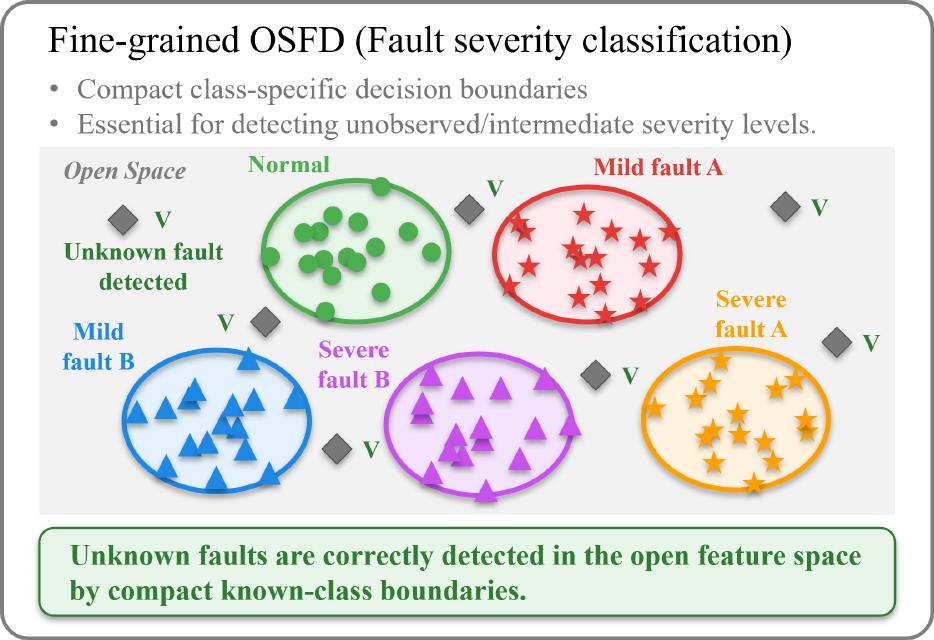}
    \caption{} 
    \label{fig:Fine_OSFD} 
\end{subfigure}
\caption{Conceptual comparison of feature representations and decision boundaries across distinct fault diagnosis paradigms. (a) Closed-set fault diagnosis failing to reject unknown anomalies. (b) Feature overlap caused by a suboptimal spectrogram representation. (c) Coarse-grained OSFD grouping continuous states into broad fault types, thereby increasing open-space risk. (d) The proposed fine-grained OSFD using compact, severity-specific boundaries to improve unknown-state rejection.}
\label{fig:OSFD_Comparison}
\end{figure}

We define the dataset as $\mathcal{D} = \{(\mathbf{x}_i, y_i)\}_{i=1}^{L}$, where $\mathbf{x}_i \in \mathbb{R}^d$ is the $i$th input feature (e.g., a preprocessed time-frequency representation), $y_i$ is its label, and $L$ is the total number of samples. We partition $\mathcal{D}$ into disjoint training and test sets, with $\mathcal{D} = \mathcal{D}_\mathrm{train} \cup \mathcal{D}_\mathrm{test}$ and $\mathcal{D}_\mathrm{train} \cap \mathcal{D}_\mathrm{test} = \emptyset$. The training set $\mathcal{D}_\mathrm{train} = \{(\mathbf{x}_i, y_i)\}_{i=1}^{N}$ contains $N$ samples whose labels belong to the set of known classes, $\mathcal{K} = \{1, 2, \dots, K\}$, where $K$ is the number of known classes. In closed-set fault diagnosis, the test set $\mathcal{D}_\mathrm{test} = \{(\mathbf{x}_i, y_i)\}_{i=N+1}^{L}$ contains only samples from $\mathcal{K}$. In OSFD, however, the test set includes samples from both the known classes $\mathcal{K}$ and an unknown class space $\mathcal{U}$. Therefore, a test label satisfies $y_i \in \mathcal{K} \cup \mathcal{U}$.

A key limitation of conventional closed-set classification arises when the softmax output is used without an explicit rejection mechanism. Softmax maps the output logits to a normalized probability distribution over the $K$ known classes, such that $\sum_{k=1}^{K}P(y=k\mid\mathbf{x})=1$,
where $P(y=k\mid\mathbf{x})$ is the predicted probability that input $\mathbf{x}$ belongs to class $k$. Figure~\ref{fig:CSFD} shows that a closed-set classifier must assign an unknown input to one of the known classes, even when the input lies outside the training distribution. Such forced classification can be particularly problematic in safety-critical applications.

Reconstruction-based OSFD identifies unknown samples by evaluating how accurately an AE trained on known classes can reconstruct them~\citep{chao2021implicit}. The AE learns a low-dimensional representation of the known-data manifold and generally produces larger reconstruction errors for samples outside this manifold. However, a single global AE represents all known classes in a shared latent space. When these classes follow distinct distributions, their representations may become entangled, creating interpolated regions in which unknown samples are reconstructed accurately. This overgeneralization reduces inter-class separability and weakens unknown-sample rejection. 

A bank of class-specific autoencoders (CSAEs) addresses this limitation by learning a separate latent manifold for each known class~\citep{huang2022class, kim2025towards}. Each CSAE specializes in reconstructing samples from one class, which increases the reconstruction error for samples that do not conform to the corresponding class-specific distribution.

\begin{problem}[Problem 1: Discriminative incapacity of a global AE] \label{prob1} Conventional reconstruction-based approaches rely on a single global AE that minimizes the overall reconstruction error. Reconstructing all known patterns simultaneously, such a network cannot enforce distinct class-specific representations, and the resulting feature space becomes entangled---leaving no clean decision boundary for rejecting unknown anomalies. \end{problem}

Most existing OSFD work classifies at the level of macroscopic fault types, where the known classes are mechanical failure modes (e.g., inner race, outer race, and ball) and unknown samples are treated as unseen fault types. This is appropriate when fault location is the diagnostic target, since different locations produce distinct characteristic frequencies and well-separated vibration signatures. Predictive maintenance, however, requires more than fault-type identification. Mechanical faults evolve gradually from incipient defects to severe damage, and type-level labels collapse this progression onto a single category. Two samples may share the same inner-race fault type but differ markedly in defect severity; grouping them hides the degradation trajectory needed for maintenance scheduling and failure prevention.

We therefore study fine-grained severity-based OSFD, which treats each degradation level within a fault type as a distinct class. The diagnostic model classifies known severities and rejects unobserved ones, providing a view of both fault type and progression.

\begin{problem}[Problem 2: Coarse-grained OSFD in Traditional Paradigms] \label{prob2} Existing OSFD methods target unseen fault types only, ignoring the continuous nature of mechanical degradation. Grouping all severities under one label is adequate for type identification but inadequate for severity tracking (Fig.~\ref{fig:Coarse_OSFD}). A severity-based paradigm (Fig.~\ref{fig:Fine_OSFD}) is needed because it captures intermediate states and yields compact, class-specific decision boundaries, reducing open-space risk. \end{problem}

Severity diagnosis requires resolving finer distinctions than type diagnosis. Adjacent severity levels share the same defect mechanism and differ only in localized spectral energy, transient impact intensity, harmonic modulation, or time-varying patterns. If these subtleties are not preserved in the feature representation, adjacent severities overlap in feature space, raising the risk of accepting intermediate degradation
states as known classes. Compact severity-specific representations and strict boundary modeling are therefore required. Because these differences are localized and non-stationary, diagnostic performance hinges on whether the input representation preserves them. This motivates the time-frequency configuration choice discussed in the following section.


\subsection{Silhouette score for representation separability evaluation} \label{sec:Silhouette}

The Silhouette score~\citep{rousseeuw1987silhouettes} measures intra-cluster compactness and inter-cluster separation. In this study, the Silhouette score is not introduced as a new clustering metric. Instead, it is used as a data-centric proxy criterion to assess whether a candidate spectrogram representation preserves the class-wise compactness and separability of fault severity states.

For a candidate STFT configuration, Let $\mathbf{S}\in\mathbb{R}^{M\times B}$ denote the 2D spectrogram matrix, where $M$ and $B$ are the numbers of time steps and frequency bins, respectively. Let $\mathbf{s}_i = \mathrm{vec}(\mathbf{S}_i) \in \mathbb{R}^{MB}$ denote the flattened spectrogram of sample $i$, and $y_i \in \mathcal{K}$ be its fault label. In the Silhouette-score computation, we treat each known fault class as a label-defined cluster, $\mathcal{C}_k=\{j \mid y_j=k\}$ for $k\in\mathcal{K}$. Thus, the cluster associated with sample $i$ is $\mathcal{C}_{y_i}$. Given a dissimilarity measure $d(\mathbf{s}_i, \mathbf{s}_j)$, we define the intra-cluster dissimilarity $a(i)$ and the nearest inter-cluster dissimilarity $b(i)$ as
\begin{align}
    a(i) = \frac{1}{|\mathcal{C}_{y_i}|-1}
    \sum_{\substack{j \in \mathcal{C}_{y_i} \\ j \neq i}}
    d(\mathbf{s}_i,\mathbf{s}_j), \quad
    b(i) = \min_{\substack{k \in \mathcal{K},\, k \ne y_i}}
    \frac{1}{|\mathcal{C}_k|}
    \sum_{j \in \mathcal{C}_k}
    d(\mathbf{s}_i,\mathbf{s}_j).
\end{align}

The Silhouette score for $N$ training samples is
\begin{align}
    \mathrm{Sil} =
    \frac{1}{N}
    \sum_{i=1}^{N} \frac{b(i)-a(i)}{\max\{a(i),b(i)\}},
    \quad \mathrm{Sil}\in[-1,1].
\end{align}
Larger positive values indicate well-separated clusters, values near zero indicate overlapping or ambiguous cluster boundaries, and negative values suggest that samples may be closer to another cluster than to their assigned cluster.

\begin{problem}[Problem 3: The STFT resolution trade-off] \label{prob3} STFT diagnostic performance is bounded by the Gabor uncertainty inequality. Suboptimal time-frequency configurations distort the non-stationary features of vibration signals and fail to preserve compact intra-class structure and inter-class margins (Fig.~\ref{fig:Suboptimal}). A data-centric parameter-selection strategy is therefore required. \end{problem}

This property is well suited to fine-grained OSFD, where adjacent severity levels produce similar vibration patterns. If the STFT does not preserve the subtle severity-dependent spectral differences, severities overlap in feature space before the diagnostic model training. Ranking candidate spectrograms by Silhouette score therefore identifies an effective STFT configuration without exhaustive network training. Section~\ref{sec:Method} addresses these three problems with an MGDC configuration search strategy and a fine-grained CSAE architecture.


\section{The proposed fine-grained open-set fault diagnosis method} \label{sec:Method}

This section presents the proposed fine-grained OSFD method, which targets Problems~\ref{prob1}--\ref{prob3}. Section~\ref{sec:Problem definition} specifies the severity-based OSFD setting. Section~\ref{sec:Overall framework} describes the overall architecture and data pipeline. Section~\ref{sec:MGDC} introduces the MGDC configuration search strategy, which selects an STFT configuration by Silhouette score. Section~\ref{sec:Architecture} details the network architecture, comprising a feature extractor and a bank of CSAEs. Section~\ref{sec:AD} formulates the dual-criteria anomaly rejection rule.


\subsection{Problem definition} \label{sec:Problem definition}

We apply the OSFD formulation in Section~\ref{sec:Fine-grained} to fault-severity diagnosis. In type-based OSFD, the known class set $\mathcal{K} = \{1, \dots, K\}$ comprises distinct fault types. In severity-based OSFD, each class instead represents an observed degradation level within a fault type, such as mild, moderate, severe damage. The unknown class set $\mathcal{U}$ contains degradation states not observed during training. The objective is to learn a diagnostic mapping $F: \mathbf{x} \rightarrow \hat{y}$ that assigns an input $\mathbf{x}$ to a known severity class when $\hat{y}\in\mathcal{K}$ and rejects it as an unknown severity state when $\hat{y}\in\mathcal{U}$.


\subsection{Overall framework of the proposed method} \label{sec:Overall framework}

\begin{figure}[ht!]
\centering 
\includegraphics[width=0.99\linewidth]{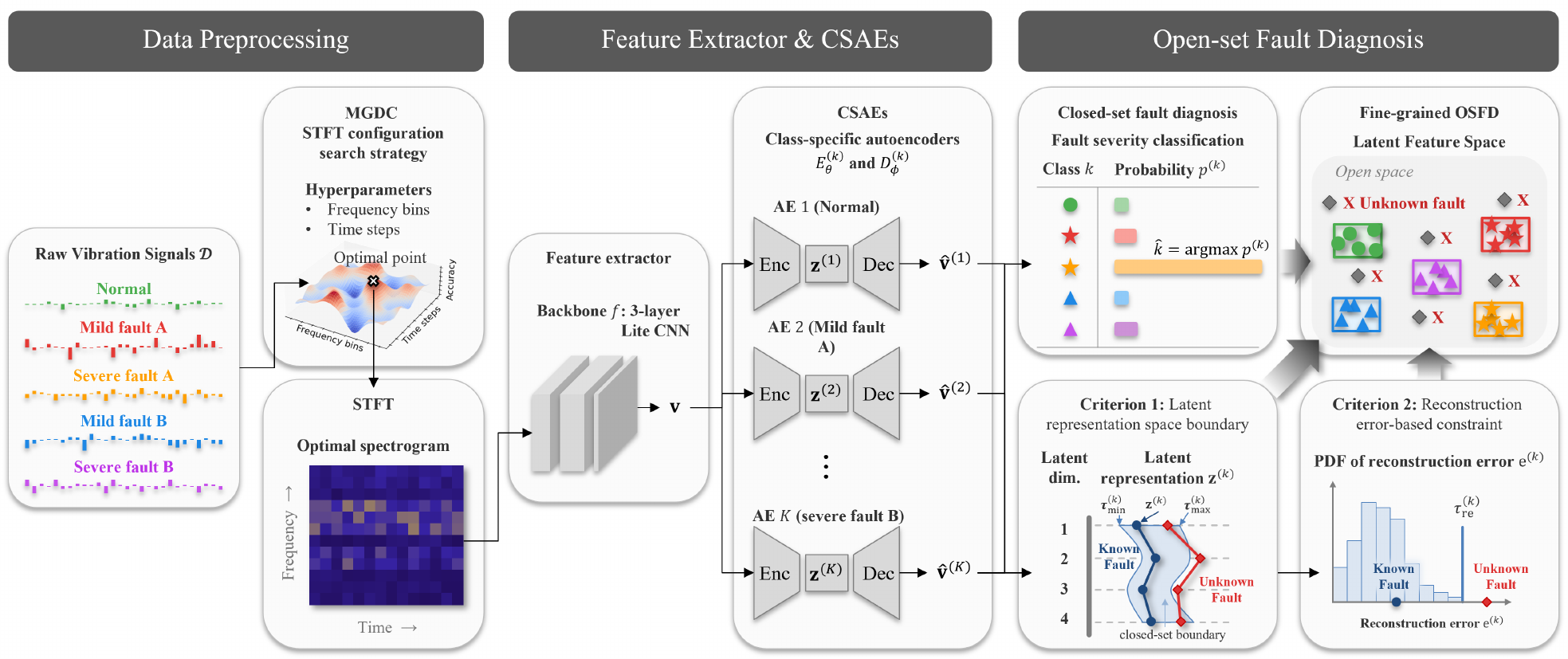} 
\caption{Overall architecture of the proposed method for fine-grained OSFD. The method comprises three sequential modules: (1) data preprocessing via the MGDC STFT configuration search, (2) representation learning using a shared feature extractor and a bank of parallel CSAEs, and (3)  known severity classification and unknown-severity rejection based on reconstruction errors and latent boundaries.} 
\label{fig:Method} 
\end{figure}

Fig.~\ref{fig:Method} provides an overview of the proposed fine-grained OSFD method for implementing the diagnostic mapping $F(\mathbf{x})$. Before describing each component in detail, we summarize the three sequential modules as follows:
\begin{itemize}
    \item \textbf{Step 1.} An MGDC search selects the STFT configuration with the highest Silhouette separability among candidates, performed once before network training (Section~\ref{sec:MGDC}).
    \item \textbf{Step 2.} The preprocessed spectrograms pass through a shared feature extractor followed by a bank of $K$ CSAEs. Each CSAE projects the features onto a latent manifold tied to one severity class, suppressing inter-class feature overlap (Section~\ref{sec:Architecture}).
    \item \textbf{Step 3.} At inference, the model first predicts a candidate severity from the minimum reconstruction error, then rejects the sample as unknown if either the reconstruction error or the latent representation falls outside the calibrated bounds of that class (Section~\ref{sec:AD}).
\end{itemize}


\subsection{Metric-guided data-centric preprocessing strategy} \label{sec:MGDC}

For each candidate STFT configuration, the raw vibration signal passes through three preprocessing stages---STFT, square-root scaling, and frequency-bin-wise normalization---before the Silhouette score is computed. The same pipeline produces the spectrograms used downstream by the feature extractor and CSAEs.

The STFT transforms the one-dimensional (1D) vibration signal into a two-dimensional (2D) time-frequency spectrogram that preserves non-stationary transients and impact signatures associated with bearing-degradation. We use the standard STFT; the contribution is the configuration selection, not a new transform. The quality of the spectrogram is limited by the STFT resolution trade-off---suboptimal choices of frequency bins and time steps distort non-stationary features and collapse class boundaries in feature space.

Two operations follow the STFT before Silhouette evaluation. The first is an element-wise square-root scaling, i.e., $\tilde{\mathbf{S}}[m,b] = \sqrt{\mathbf{S}[m,b]}$, with $m=0,\dots,M-1$ and $b=0,\dots,B-1$ indexing time frame and frequency bin, where $\mathbf{S}[m,b]\in\mathbb{R}^{M\times B}$ denotes the power spectrogram obtained from the STFT, and $M$ and $B$ are the numbers of time steps and frequency bins, respectively. The square root converts the spectrogram from power to amplitude and compresses the dynamic range so low-energy harmonics remain visible alongside high-energy impacts.

The second is a per-frequency-bin min-max normalization, which removes scale variation across operating speeds and loads. The statistics $S^{(b)}_{\min}, S^{(b)}_{\max}$ are estimated on the training set only and reused at test time to prevent leakage. For each frequency bin $b$,
\begin{align}
    S^{(b)}_{\min} &= \min_{i,m} \tilde{\mathbf{S}}_i[m,b]\quad\text{and}\quad 
    S^{(b)}_{\max} = \max_{i,m} \tilde{\mathbf{S}}_i[m,b],
\end{align}
where $i$ is the training sample index. The normalized spectrogram is 
\begin{align}
    \hat{\mathbf{S}}[m, b] = \frac{\tilde{\mathbf{S}}[m, b] - S^{(b)}_{\min}} {S^{(b)}_{\max} - S^{(b)}_{\min}}.
\end{align}
Normalizing each frequency bin individually decouples the learned representation from absolute spectral magnitudes and improves train/test distribution consistency.

Let $\mathcal{H} = \{h_j\}_{j=1}^J$ be the search space of candidate STFT configurations, with $J$ candidates each specifying one time-frequency resolution. For every $h_j$, the training signals pass through the same three-stage pipeline and yield normalized spectrograms $\hat{\mathbf{S}}^{(j)}$. These are flattened into 1D vectors, the known severity labels act as cluster assignments, and the Silhouette score $\mathrm{Sil}(h_j)$ is computed. This avoids trial-and-error and exhaustive network training over all candidates.
A higher $\mathrm{Sil}(h_j)$ indicates greater intra-class compactness and clearer inter-class separation for configuration $h_j$. We sort the $J$ candidates by $\mathrm{Sil}(h_j)$ in descending order and retain the top-$R$ as a reduced candidate set $\mathcal{H}_{R} \subset \mathcal{H}$ with $|\mathcal{H}_{R}| = R$.

The final configuration is not chosen by the maximum Silhouette score alone. The score reduces the search space, and diagnostic validation on $\mathcal{H}_{R}$ then selects
\begin{align}
    h^{*} = \underset{h_j \in \mathcal{H}_{R}}{\arg\max}\; H_{\mathrm{eval}}(h_j),
\end{align} where $H_{\mathrm{eval}}(h_j)$ is the H-score on the selection set under configuration $h_j$.

Unlike network-in-the-loop HPO methods (TPE, Hyperband, BOHB), MGDC does not train the diagnostic network for every STFT candidate. The training-free Silhouette score screens $\mathcal{H}$, and network-based validation is restricted to the high-scoring subset $\mathcal{H}_{R}$. Sections~\ref{sec:STFT_C} and~\ref{sec:STFT_P} show that the Silhouette ranking includes near-optimal configurations within the top few candidates. The selected configuration produces the final normalized spectrogram $\hat{\mathbf{S}}^{*}$, which feeds the feature extractor and CSAE modules. Further details are in Appendix~\ref{apdx:MGDC}.

    
\subsection{Proposed network architecture} \label{sec:Architecture}

The diagnostic network combines a shared feature extractor with a
parallel bank of CSAEs, designed to construct compact and isolated latent manifolds for each degradation stage. 
Using the selected optimal STFT configuration $h^*$, the training and test sets in the spectrogram domain are 
\begin{align}
    \mathcal{D}_{\mathrm{train}}^\mathrm{S} &= \{(\hat{\mathbf{S}}_{\ell}^{*}, \tilde{y}_{\ell})\}_{\ell=1}^{N_\mathrm{S}}\quad\text{and}\quad \mathcal{D}_{\mathrm{test}}^\mathrm{S} = \{(\hat{\mathbf{S}}_{\ell}^{*}, \tilde{y}_{\ell})\}_{\ell=N_\mathrm{S}+1}^{L_\mathrm{S}},
\end{align}
where $\hat{\mathbf{S}}_{\ell}^{*}$ is the $\ell$th normalized spectrogram under $h^*$, $\tilde{y}_{\ell}$ its fault label, $N_\mathrm{S}$ the number of training spectrograms, and $L_\mathrm{S}$ the total dataset size.


\subsubsection{Feature extractor} \label{sec:Feature Extractor}

Large architectures such as Vision Transformers~\citep{xu2023devit} carry substantial compute overhead, making real-time deployment on resource-constrained edge devices impractical. We therefore use a 3-layer lightweight CNN~\citep{mukherjee2021light} as the feature extractor $f(\cdot)$, following shallow-CNN designs used in industrial fault diagnosis.
The backbone is kept intentionally small to isolate the contribution of the proposed CSAE bank and dual-criteria rejection. Improved H-score performance therefore reflects these design choices rather than model capacity or network depth.

The extractor passes the 2D spectrogram $\hat{\mathbf{S}}^{*}$ through three convolutional blocks that extract hierarchical spatial features. The resulting maps are flattened into a 1D feature vector $\mathbf{v} = f(\hat{\mathbf{S}}^{*}) \in \mathbb{R}^{d_v}$, which is routed to the parallel CSAE bank for severity-specific analysis.


\subsubsection{Class-specific autoencoders} \label{sec:CSAE}

To prevent latent entanglement across severity levels, we use a parallel
bank of $K$ independent CSAEs. The $k$th CSAE has a fully connected (FC)
encoder $E_\theta^{(k)}$ and decoder $D_\phi^{(k)}$.

On the forward pass, the feature vector $\mathbf{v} \in \mathbb{R}^{d_v}$ is routed to all $K$ CSAEs in parallel. The $k$th encoder compresses $\mathbf{v}$ into a latent representation $\mathbf{z}^{(k)} = E_\theta^{(k)}(\mathbf{v}) \in \mathbb{R}^{d_z}$ with $d_z < d_v$, and the $k$th decoder reconstructs it as $\hat{\mathbf{v}}^{(k)} = D_\phi^{(k)}(\mathbf{z}^{(k)})$. The class-specific reconstruction error is $e^{(k)}(\mathbf{v}) = \|\mathbf{v} - \hat{\mathbf{v}}^{(k)}\|_1$. We use the $L_1$ norm rather than squared $L_2$ because $L_1$ penalizes errors linearly rather than quadratically, providing robustness to the impulsive transients typical of bearing fault signatures.

The negative reconstruction error acts as a class-affinity score, and the softmax-normalized class probability is
\begin{align}
    p^{(k)} = \frac{\exp(-e^{(k)})}{\sum_{t=1}^{K} \exp(-e^{(t)})},
\end{align}
so a smaller $e^{(k)}$ yields a larger $p^{(k)}$. Because $p^{(k)}$ is defined from the negative reconstruction errors, the cross-entropy loss
\begin{align}
    \mathcal{L}_\mathrm{CE}
    = -\sum_{k=1}^{K} \mathbb{I}(y = k)\, \log p^{(k)},
\end{align}
drives the CSAE matched to the ground-truth class toward the lowest reconstruction error among the $K$ branches. The indicator $\mathbb{I}(y=k) \in \{0,1\}$ equals the $k$th element of the one-hot encoding of $y$. The feature extractor and all $K$ CSAEs are trained jointly by minimizing $\mathcal{L}_\mathrm{CE}$. Backpropagation drives the extractor toward severity-discriminative features and each CSAE toward accurate reconstruction of its own class.


\subsubsection{Reconstruction error-based diagnostic decision} \label{sec:Diagnostic decision}

\begin{figure}[ht!]
\centering
\includegraphics[width=0.8\linewidth]{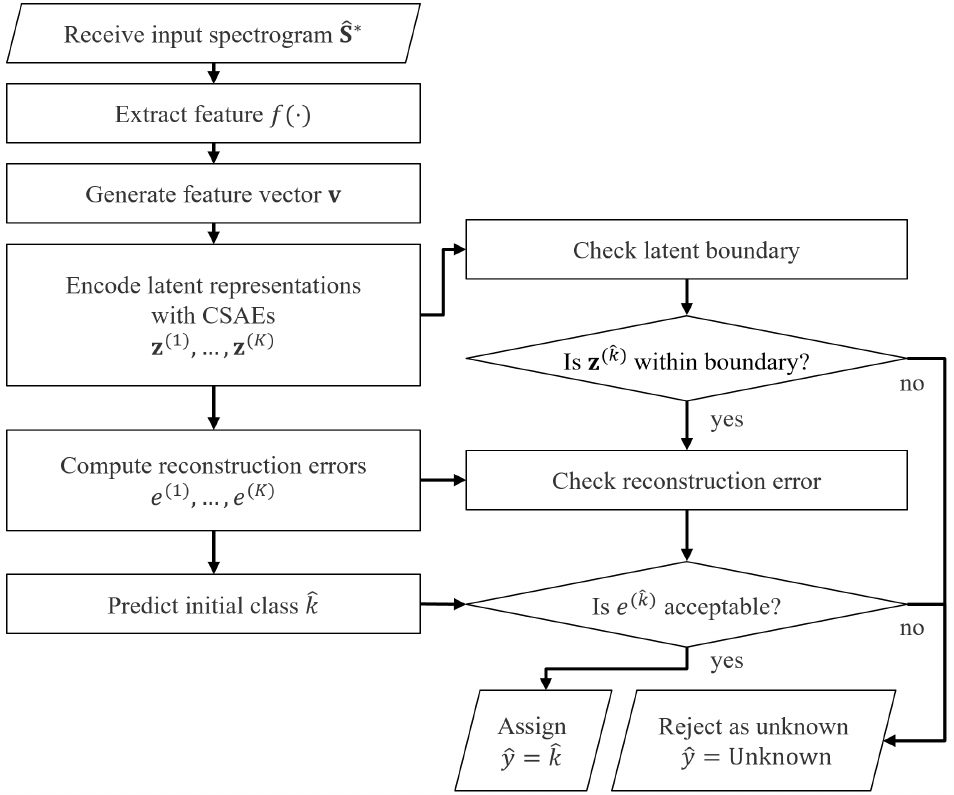}
\caption{Inference flow of the proposed OSFD framework. The input spectrogram $\hat{\mathbf{S}}^{*}$ is converted into a feature vector and evaluated using class-specific autoencoders. The initial class prediction $\hat{k}$ is accepted as a known class only if both the latent boundary and reconstruction error criteria are satisfied; otherwise, the sample is classified as unknown.} 
\label{fig:Flowchart} 
\end{figure}

At inference, the input is assigned to the known severity class whose CSAE produces the smallest reconstruction error, equivalently the largest softmax probability:
\begin{align}
    \hat{k} = \underset{k \in \mathcal{K}}{\arg\max}~p^{(k)}.
\end{align}
Because both known-class prediction and anomaly detection use class-specific reconstruction information, the classification output directly supports the subsequent open-set rejection step. 

Fig.~\ref{fig:Flowchart} shows the complete inference and decision-making pipeline of the proposed method. The normalized input spectrogram $\hat{\mathbf{S}}^{*}$ is passed through the shared feature extractor $f(\cdot)$ to obtain the feature vector $\mathbf{v}=f(\hat{\mathbf{S}}^{*})$. Then $\mathbf{v}$ is fed to all $K$ CSAEs in parallel, producing the latent representations $\{\mathbf{z}^{(k)}\}_{k=1}^{K}$ and the corresponding reconstruction errors $\{e^{(k)}\}_{k=1}^{K}$. The class with the smallest reconstruction error is selected as the initial known-class candidate $\hat{k}$. Finally, the candidate label is accepted as $\hat{y}=\hat{k}$ only when both criteria using latent and reconstruction error are satisfied; otherwise, the input is assigned to the unknown class.
The following section defines the dual-criteria anomaly rejection rule and the final diagnostic output $\hat{y}$.


\subsection{Anomaly rejection mechanism} \label{sec:AD}

The closed-set predictor returns the most likely known class $\hat{k}$ even when the input is anomalous, because softmax confidence does not distinguish in-distribution from out-of-distribution inputs. We verify the candidate $\hat{k}$ with a dual-criteria rejection rule based on (1) the latent representation and (2) the reconstruction error.


\subsubsection{Criterion 1: Latent representation space boundary} \label{sec:la}

The first criterion evaluates whether the latent representation of the test sample lies within the training distribution of class $\hat{k}$. Distance-based methods use a distance metric (Euclidean, Mahalanobis) to bound the latent space with a hypersphere or ellipsoid. However, these distance-based approaches inherently compress the high-dimensional latent vector into a single scalar value. This compression masks fine-grained, directional fault signatures that appear only in specific frequency bands or latent dimensions, thereby reducing the separability between known and unknown fault states in the latent space. Thus, unknown anomalies can exhibit deceptively normal distance scores, leading to unreliable open-set rejection.

We instead define multi-dimensional boundaries for each class $k$ during the training phase. Let $\mathcal{Z}^{(k)}=\{\mathbf{z}_\ell^{(k)} \mid \ell=1,\dots,N_\mathrm{S}, \tilde{y}_\ell=k \}$ be the latent representations of training samples in class $k$, where $\tilde{y}_\ell$ is the true label of the $\ell$th normalized spectrogram. Instead of relying on a single scalar distance threshold, we construct dimension-wise bounds for each latent coordinate using the class-specific training distribution, without imposing a statistical independence assumption. This criterion is intended to detect coordinate-wise deviations that may be hidden in a scalar distance score. The element-wise lower and upper bounds are 
\begin{align}
    \boldsymbol{\tau}_{\min}^{(k)} &= \mathcal{Q}_{1-\alpha}(\mathcal{Z}^{(k)}) \in \mathbb{R}^{d_z}\quad\text{and}\quad
    \boldsymbol{\tau}_{\max}^{(k)} = \mathcal{Q}_{\alpha}(\mathcal{Z}^{(k)}) \in \mathbb{R}^{d_z},
\end{align}
where $\mathcal{Q}_{\alpha}(\cdot)$ is the element-wise $\alpha$-quantile across the training samples and $\alpha$ is the target confidence percentile. With $\alpha=0.9999$, the marginal false-rejection rate is $2(1-\alpha)=2\times 10^{-4}$. By the union bound, the joint false-rejection rate is at most $2 d_z (1-\alpha)$, i.e., $4 \times 10^{-4}$ ($0.04\%$) at $d_z = 2$. The sensitivity of $\alpha$ is analyzed in Section~\ref{sec:Threshold}. A test sample is flagged if $\mathbf{z}_\ell^{(\hat{k})}$ falls outside $[\boldsymbol{\tau}_{\min}^{(\hat{k})}, \boldsymbol{\tau}_{\max}^{(\hat{k})}]$ on even one coordinate.


\subsubsection{Criterion 2: Reconstruction error-based constraint} \label{sec:re}
The second criterion catches anomalies that slip through Criterion~1. Even if an unknown input's latent coordinates fall inside the bounds for class $\hat{k}$, the decoder $D_\phi^{(k)}$---trained only on class $\hat{k}$ patterns---reconstructs the input as a class $\hat{k}$ exemplar. The resulting reconstruction error $e^{(\hat{k})}(\mathbf{v})$ is large, exposing the anomaly.

Let $\mathcal{E}^{(k)} = \{e_\ell^{(k)} \mid \ell=1,\dots,N_{\mathrm{S}},\, \tilde{y}_\ell = k\}$ be the reconstruction errors of training samples in class~$k$. The class-specific reconstruction threshold is the same $\alpha$-quantile,
\begin{align}
    \tau_{\mathrm{re}}^{(k)} = \mathcal{Q}_{\alpha}(\mathcal{E}^{(k)}) \in \mathbb{R}.
\end{align}
A test sample is flagged as a feature-level anomaly if $e_\ell^{(\hat{k})} > \tau_{\mathrm{re}}^{(\hat{k})}$.


\subsubsection{Final diagnostic decision logic} 
To robustly prevent unknown anomalies from being incorrectly accepted as known severity classes, the proposed method uses a strict logical conjunction (AND-to-accept) between the two criteria. A test sample is accepted as the known severity $\hat{k}$ only if it simultaneously satisfies both the latent-based and the reconstruction error-based criteria. Otherwise, the sample is rejected as an unknown anomaly. This conservative decision strategy prioritizes rejection reliability by reducing false acceptance of unknown faults, which is particularly important in safety-critical industrial diagnostic environments.
    
The final diagnostic output $\hat{y}=F(\mathbf{x})$ is finalized as the following piecewise decision rule:
\begin{align}
    \hat{y} = 
    \begin{cases} 
        \hat{k}, & \text{if } \left( \boldsymbol{\tau}_{min}^{(\hat{k})} \le \mathbf{z}^{(\hat{k})} \le \boldsymbol{\tau}_{max}^{(\hat{k})} \right) \land \left( e^{(\hat{k})} \le \tau_{re}^{(\hat{k})} \right), \\
        \text{Unknown}, & \text{otherwise}.
    \end{cases}
\end{align}
Note that the inequality for the latent vector $\mathbf{z}^{(\hat{k})}$ is applied element-wise across all dimensions.

Algorithm~\ref{alg:dualcriteria} provides the complete inference and anomaly rejection procedure of the proposed CSAE-based OSFD method. The inputs to the algorithm are the STFT domain training set $\mathcal{D}_{\mathrm{train}}^{\mathrm{S}}$, the STFT domain test set $\mathcal{D}_{\mathrm{test}}^{\mathrm{S}}$, the trained feature extractor $f(\cdot)$, the trained bank of CSAEs $\{(E_\theta^{(k)},D_\phi^{(k)})\}_{k=1}^{K}$, and the confidence percentile $\alpha$. The training set is used only to calibrate the class-specific latent boundaries and reconstruction-error thresholds from known-class samples. The algorithm then processes each test sample without access to its ground-truth label. It first identifies the candidate known severity class from the class-specific reconstruction errors and then applies the two rejection criteria to accept the candidate or label the sample as unknown. The output is the set of predicted diagnostic labels $\{\hat{y}_{\ell}\}_{\ell=N_{\mathrm{S}}+1}^{L_{\mathrm{S}}}$ for all test samples. This algorithm specifies the calibration, classification, and rejection steps required to reproduce the proposed inference procedure.

\label{sec:AD_logic}
\begin{algorithm}[H]
\caption{Open-set inference with CSAEs}
\label{alg:dualcriteria}
\begin{algorithmic}
\Require STFT-domain training and test set $\mathcal{D}_{\mathrm{train}}^{\mathrm{S}}$, $\mathcal{D}_{\mathrm{test}}^{\mathrm{S}}$, trained feature extractor $f(\cdot)$, trained CSAEs $\{(E_\theta^{(k)},D_\phi^{(k)})\}_{k=1}^{K}$, and confidence percentile $\alpha$
\Ensure Predicted labels $\{\hat{y}_{\ell}\}_{\ell=N_\mathrm{S}+1}^{L_\mathrm{S}}$
\end{algorithmic}

\noindent
\begin{minipage}[t]{0.48\linewidth}
\begin{algorithmic}
\Statex \textbf{Stage 1: Threshold calibration}

\For{$k \in \{1,\dots,K\}$}
    \State Initialize $\mathcal{Z}^{(k)} \leftarrow \emptyset$ and $\mathcal{E}^{(k)} \leftarrow \emptyset$

    \For{$(\hat{\mathbf{S}}_{\ell}^{*},\tilde{y}_{\ell}) \in \mathcal{D}_{\mathrm{train}}^{\mathrm{S}}$ with $\tilde{y}_{\ell}=k$}
        \State Extract: $\mathbf{v}_{\ell}=f(\hat{\mathbf{S}}_{\ell}^{*})$
        \State Encode: $\mathbf{z}_{\ell}^{(k)}=E_{\theta}^{(k)}(\mathbf{v}_{\ell})$
        \State Reconstruct: $\hat{\mathbf{v}}_{\ell}^{(k)}=D_{\phi}^{(k)}(\mathbf{z}_{\ell}^{(k)})$
        \State Compute: $e_{\ell}^{(k)}=\left\lVert \mathbf{v}_{\ell}-\hat{\mathbf{v}}_{\ell}^{(k)} \right\rVert_{1}$
        \State Update $\mathcal{Z}^{(k)}\leftarrow\mathcal{Z}^{(k)}\cup\{\mathbf{z}_{\ell}^{(k)}\}$
        \State Update $\mathcal{E}^{(k)}\leftarrow\mathcal{E}^{(k)}\cup\{e_{\ell}^{(k)}\}$
    \EndFor

    \State Estimate:
    \Statex $\boldsymbol{\tau}_{\min}^{(k)}=\mathcal{Q}_{1-\alpha}(\mathcal{Z}^{(k)}),\quad \boldsymbol{\tau}_{\max}^{(k)}=\mathcal{Q}_{\alpha}(\mathcal{Z}^{(k)})$
    \State Estimate:
    \Statex $\tau_{\mathrm{re}}^{(k)}=\mathcal{Q}_{\alpha}(\mathcal{E}^{(k)})$
\EndFor
\end{algorithmic}
\end{minipage}
\hfill
\begin{minipage}[t]{0.48\linewidth}
\begin{algorithmic}
\Statex \textbf{Stage 2: Open-set inference}

\For{$\hat{\mathbf{S}}_{\ell}^{*} \in \mathcal{D}_{\mathrm{test}}^{\mathrm{S}}$}
    \State Extract: $\mathbf{v}_{\ell} = f(\hat{\mathbf{S}}_{\ell}^{*})$
    
    \For{$k \in \{1,\dots,K\}$}
        \State Encode: $\mathbf{z}_{\ell}^{(k)} = E_{\theta}^{(k)}(\mathbf{v}_{\ell})$
        \State Reconstruct: $\hat{\mathbf{v}}_{\ell}^{(k)} = D_{\phi}^{(k)}(\mathbf{z}_{\ell}^{(k)})$
        \State Compute: $e_{\ell}^{(k)} = \left \lVert \mathbf{v}_{\ell}-\hat{\mathbf{v}}_{\ell}^{(k)} \right \rVert_{1}$
    \EndFor
    
    \State Obtain:
    \Statex $\hat{k}_{\ell} = \underset{k}{\arg\min} e_{\ell}^{(k)}$

    \If{$\boldsymbol{\tau}_{\min}^{(\hat{k}_{\ell})} \leq \mathbf{z}_{\ell}^{(\hat{k}_{\ell})} \leq \boldsymbol{\tau}_{\max}^{(\hat{k}_{\ell})}$ and $e_{\ell}^{(\hat{k}_{\ell})} \leq \tau_{\mathrm{re}}^{(\hat{k}_{\ell})}$}
        \State $\hat{y}_{\ell} = \hat{k}_{\ell}$
    \Else
        \State $\hat{y}_{\ell} = \mathrm{Unknown}$
    \EndIf
\EndFor
\State \Return $\{\hat{y}_{\ell}\}_{\ell=N_\mathrm{S}+1}^{L_\mathrm{S}}$
\end{algorithmic}
\end{minipage}
\end{algorithm}


\section{Results} \label{sec:Results}

This section evaluates the proposed fine-grained OSFD framework through numerical experiments. We begin by describing the common experimental setup in Section~\ref{sec:Experimental_setup}, including the hardware environment, training configuration, and dataset split. Section~\ref{sec:Evaluation_metrics} then defines the evaluation metrics used to assess both known-class classification and unknown-class rejection. Section~\ref{sec:Architecture_hyperparameter} specifies the model architecture and hyperparameter settings, and  Section~\ref{sec:Baseline} summarizes the baseline methods used for comparison. Sections~\ref{sec:CWRU} and \ref{sec:PU} present the results on the Case Western Reserve University (CWRU)~\citep{smith2015rolling} and Paderborn University (PU)~\citep{lessmeier2016condition} bearing datasets, respectively. For each dataset, we describe the dataset configuration, select the STFT configuration, and compare performance with baseline methods. Finally, Section~\ref{sec:Cost} discusses the computational cost of the proposed method.


\subsection{Experimental setup} \label{sec:Experimental_setup}

All experiments were conducted on a workstation equipped with an Intel(R) Xeon(R) w3-2435 CPU and an NVIDIA RTX A5000 GPU with 24GB of memory. The network was optimized using the Adam optimizer with a learning rate of $0.0001$. The batch size was fixed at $25$ for all experimental cases, and the number of training epochs was set to $10$. The dataset was divided into training, validation, selection, and test sets with split ratios of $0.6$, $0.1$, $0.1$, and $0.2$, respectively.  The training set was used to optimize the network parameters, and the validation set was used to monitor the training process. The selection set was used only for model-selection procedures, including STFT configuration selection, CSAE architecture hyperparameter selection, and confidence percentile threshold selection. The test set was kept independent of training procedures and was used for final performance evaluation.


\subsection{Evaluation metrics} \label{sec:Evaluation_metrics}

We use a comprehensive set of evaluation metrics to assess the diagnostic performance and OSFD capability of the proposed method. In the OSFD setting, the model must maintain high classification accuracy for known fault classes while reliably detecting samples from unknown fault classes. To evaluate both diagnostic precision and detection sensitivity, we report the micro F1-score, macro F1-score, unknown detection accuracy (UDA), closed-set accuracy (CSA), open-set accuracy (OSA), and harmonic score (H-score).  Appendix~\ref{apdx:Detailed OSFD evaluation metrics} provides the detailed mathematical definitions of these metrics.

For each open-set task, we compute UDA, CSA, OSA, F1-score, and H-score independently. Tables~\ref{tab:Result_C} and~\ref{tab:Result_P} report the values averaged over all tasks within each diagnostic granularity, i.e., fault type or fault severity. Therefore, the reported H-score represents the mean of the task-wise H-scores and is not necessarily identical to the H-score computed from the averaged UDA and CSA values.


\subsection{Model architecture hyperparameter specifications} \label{sec:Architecture_hyperparameter}

\begin{table}[ht!]
\centering 
\caption{Comprehensive architectural configuration of the proposed diagnostic method. The model consists of a shared feature extractor and a parallel bank of $K$ class-specific autoencoders. The feature extractor uses hierarchical 2D convolutions to transform spectrogram inputs into a feature vector $\mathbf{v}$. Each CSAE branch, optimized for a specific severity class, then performs manifold learning through a $d_z$-dimensional latent representation to isolate class-specific structural signatures. The $K \times$ notation denotes the parallel processing units corresponding to the known severity levels, followed by error-based softmax classification over the negative reconstruction errors.} 
\label{tab:Model}
\renewcommand{\arraystretch}{1.2}
\begin{tabular}{cccccc} 
\toprule
    \multicolumn{2}{c}{\textbf{Component}} & \textbf{Layer} & \textbf{Configuration} & \textbf{Activation} & \textbf{Output size} \\ 
\midrule
    \multicolumn{2}{c}{Input} & - & - & - & $[1, F, T]$ \\ 
\midrule
    \multirow{4}{*}{\begin{tabular}[c]{@{}c@{}}Feature \\ extractor\end{tabular}} & \multirow{4}{*}{\begin{tabular}[c]{@{}c@{}}Lightweight \\ CNN\end{tabular} } 
    & Conv2d 1 & $(3\times3, 1, 1)^*$ & LeakyReLU(0.2) & $[32, F, T]$  \\ 
    & & Conv2d 2 & $(1\times1, 1, 0)^*$ & LeakyReLU(0.2) & $[64, F, T]$  \\ 
    & & Conv2d 3 & $(1\times1, 1, 0)^*$ & LeakyReLU(0.2) & $[1, F, T]$ \\ 
    & & Flatten   & - & - & $[F \times T]$ \\ 
\midrule
    \multirow{6}{*}{CSAEs} & \multirow{2}{*}{Encoder} 
    & FC 1 & $K \times d_h$ & ReLU & $[K, d_h]$ \\ 
    & & FC 2 & $K \times d_h$ & ReLU & $[K, d_h]$ \\ 
\cmidrule{2-6}
    & Latent 
    & FC 3 & $K \times d_z$ & Linear & $[K, d_z]$\\ 
\cmidrule{2-6}
    & \multirow{3}{*}{Decoder} 
    & FC 4 & $K \times d_h$ & ReLU & $[K, d_h]$ \\ 
    & & FC 5 & $K \times d_h$ & ReLU & $[K, d_h]$ \\ 
    & & FC 6 & $K \times (F \times T)$ & Linear & $[K, F \times T]$ \\ 
\midrule
    \multicolumn{2}{c}{Error-based classifier} & Softmax & $K$ & - & $[K]$ \\ 
\bottomrule
    \multicolumn{6}{l}{\footnotesize *Note that the configuration format for Conv2d is (kernel size, stride, padding).} \\
\end{tabular}
\end{table}

Table~\ref{tab:Model} presents the layer-by-layer configuration of the proposed method, including the fixed feature extractor and the optimized CSAE module. The network takes a preprocessed spectrogram with $F$ frequency bins and $T$ time steps as input. The feature extractor first processes this two-dimensional input through three convolutional layers, each followed by a LeakyReLU activation with a coefficient of $0.2$, to extract high-level representations. The final convolutional output is then flattened into a feature vector $v \in \mathbb{R}^{d_v}$, where $d_v = F \times T$. This one-dimensional feature vector is subsequently fed into $K$ parallel CSAE branches.

Within each CSAE branch, the encoder uses two fully connected (FC) layers to compress the feature vector into a $d_h$-dimensional hidden representation and then maps it to a $d_z$-dimensional latent representation. We use a linear activation at the latent layer to preserve the structural manifold of the extracted features without additional nonlinear distortion. The decoder then uses two FC layers to expand the latent representation back to $d_v$ dimensions. The final FC layer outputs an $F \times T$-dimensional reconstruction vector, thereby performing feature-level reconstruction of the flattened feature representation. Finally, we apply a softmax operation to the negative reconstruction errors to obtain the probability distribution over the $K$ known fault classes.

\begin{figure}[ht!]
\centering 
\includegraphics[width=0.99\linewidth]{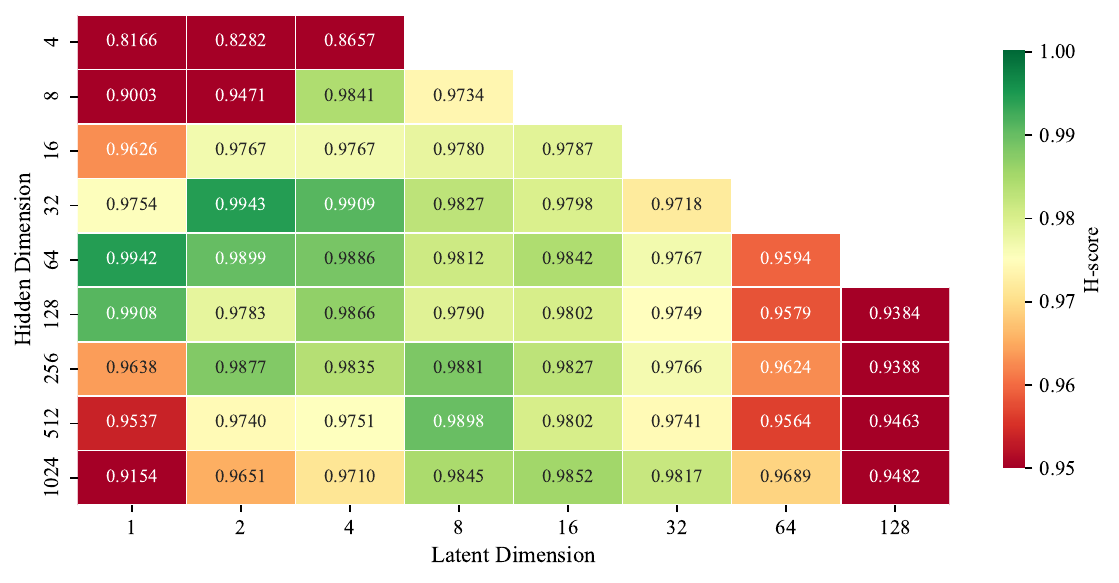}
\caption{Hyperparameter search space for the CSAE architecture.} 
\label{fig:CWRU_HPO_heatmap}
\end{figure}

The internal dimensions of the CSAEs directly affect the network's ability to compress known-class features and reject unobserved anomalies. To select suitable hidden and latent dimensions, we conducted a parametric study on the selection set over different hidden--latent dimension combinations in the CSAE architecture. We selected
the hidden and latent dimensions from $\{4, 8, 16, 32, 64, 128,$ $256, 512, 1024\}$ and $\{1, 2, 4, 8, 16, 32, 64, 128\}$, respectively. Configurations with a latent dimension larger than the hidden dimension were excluded to preserve the AE bottleneck. Figure~\ref{fig:CWRU_HPO_heatmap} shows that $d_h=32$ and $d_z=2$ achieved the highest H-score of $0.9943$; these values were therefore used in subsequent experiments. 

\begin{figure}[ht!]
\centering 
\begin{subfigure}[b]{0.48\textwidth}
    \includegraphics[width=\linewidth]{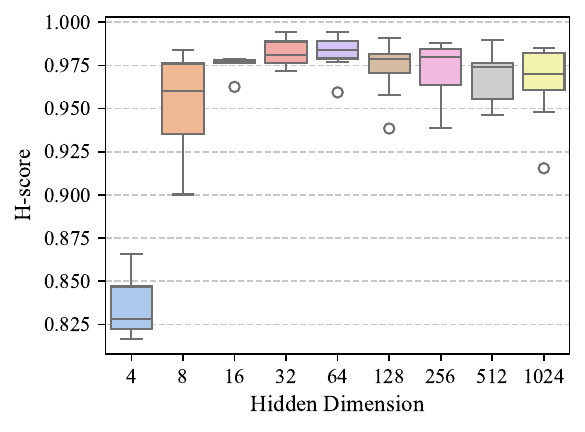}
    \caption{}
    \label{fig:CWRU_HPO_hid}
\end{subfigure}
\begin{subfigure}[b]{0.48\textwidth}
    \includegraphics[width=\linewidth]{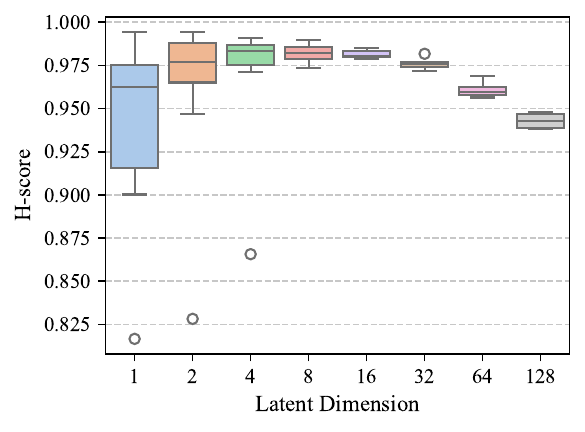}
    \caption{}
    \label{fig:CWRU_HPO_lat}
\end{subfigure}
\caption{H-score performance distributions under different (a) hidden and (b) latent dimensions.}
\label{fig:CWRU_HPO_boxplot} 
\end{figure}

Fig.~\ref{fig:CWRU_HPO_hid} shows that small hidden dimensions yield unstable H-scores because their limited representation capacity cannot adequately encode fault-related patterns in the spectrograms. This instability becomes more pronounced at small latent dimensions, where excessive compression causes substantial information loss. Fig.~\ref{fig:CWRU_HPO_lat} shows that large latent dimensions weaken the AE bottleneck, allowing unknown samples to be reconstructed with low errors. This reduces the separation between known and unknown samples and degrades anomaly rejection. Intermediate hidden and latent dimensions therefore provide a more effective balance between representation capacity and compression.


\subsection{Baseline models} \label{sec:Baseline}

We evaluate the accuracy and computational efficiency of the proposed method by comparing it with the representative open-set diagnostic baselines. We select four benchmark methods: global autoencoder (Global AE) for reconstruction-based detection, OpenMax for extreme value theory-based calibration, CPL for distance-based metric learning, and ARPL for adversarial reciprocal point learning. This selection provides a diverse and rigorous basis for assessing the proposed method against established OSFD paradigms. 

For STFT configuration selection, we compare the proposed MGDC strategy with representative HPO-based search strategies, including the Tree-structured Parzen Estimator (TPE)~\citep{bergstra2011algorithms}, Hyperband~\citep{li2018hyperband}, and Bayesian Optimization and HyperBand (BOHB)~\citep{falkner2018bohb}. We implement TPE using the TPE sampler in Optuna~\citep{akiba2019optuna}. TPE performs sequential model-based optimization by using previous trial results to sample promising configurations from the search space. Hyperband is a multi-fidelity search strategy that allocates small training budgets to many configurations and progressively retains promising candidates through successive halving. BOHB combines Bayesian optimization with Hyperband-style multi-fidelity resource allocation, thereby using model-based sampling while reducing the cost of evaluating poor configurations at early stages. Unlike these HPO-based strategies, which require network-based performance feedback through full or partial training, the proposed MGDC strategy screens candidate STFT configurations using a training-free representation-level separability metric.


\subsection{Case 1: CWRU dataset} \label{sec:CWRU}

This case study evaluates the proposed method on the Case Western Reserve University (CWRU) bearing dataset~\citep{smith2015rolling} for both fault type and fault severity OSFD. The objective is to examine whether the proposed method can maintain accurate classification of known fault classes while rejecting unseen fault states, with particular emphasis on fine-grained severity levels within the same fault type.

\subsubsection{Dataset and task description of the CWRU dataset} \label{sec:Dataset C}

\begin{table}[ht!]
\centering 
\caption{Specifications of the CWRU dataset.} 
\label{tab:Dataset C}
\renewcommand{\arraystretch}{1.2}
\begin{tabular}{ccccc}
\toprule
    \textbf{Class} & \textbf{Fault type} & \textbf{Fault severity} [inch] & \textbf{Fault severity} [mm] &\textbf{Bearing code} \\ 
\midrule
    CB1 & Ball fault       & 0.007 & 0.1778 & B007\_3 \\ 
    CB2 & Ball fault       & 0.014 & 0.3556 & B014\_3 \\ 
    CB3 & Ball fault       & 0.021 & 0.5334 & B021\_3 \\ 
    CI1 & Inner race fault & 0.007 & 0.1778 & IR007\_3 \\ 
    CI2 & Inner race fault & 0.014 & 0.3556 & IR014\_3 \\ 
    CI3 & Inner race fault & 0.021 & 0.5334 & IR021\_3 \\ 
    CN  & Normal           & -     & -      & Normal\_3 \\ 
    CO1 & Outer race fault & 0.007 & 0.1778 & OR007\_6\_3 \\ 
    CO2 & Outer race fault & 0.014 & 0.3556 & OR014\_6\_3 \\ 
    CO3 & Outer race fault & 0.021 & 0.5334 & OR021\_6\_3 \\
\bottomrule
\end{tabular}
\end{table}

We categorize the CWRU bearing dataset into ten distinct classes by jointly considering fault types and severity levels. These classes include the normal condition and nine faulty conditions defined by three fault locations and three defect diameters as shown in Table~\ref{tab:Dataset C}. For consistent evaluation, we use only vibration signals acquired under a constant motor load of 3 horsepower. 

\begin{table}[ht!]
\centering 
\caption{Configuration of OSFD tasks for the CWRU dataset. Tasks TC1--TC3 focus on unknown fault types, while TC4-TC12 address the diagnosis of unseen fault severity levels.} 
\label{tab:Task C}
\renewcommand{\arraystretch}{1.2}
\begin{tabular}{cccc}
\toprule
    \textbf{Granularity} & \textbf{Task} & \textbf{Known class} & \textbf{Unknown class} \\ 
\midrule
    \multirow{3}{*}{\begin{tabular}[c]{@{}c@{}}\textbf{Fault}\\ \textbf{type}\end{tabular}} 
    & TC1 & \phantom{CB1, CB2, CB3, }CI1, CI2, CI3, CN, CO1, CO2, CO3 & CB1, CB2, CB3 \\
    & TC2  & CB1, CB2, CB3, \phantom{CI1, CI2, CI3, }CN, CO1, CO2, CO3 & CI1, CI2, CI3 \\
    & TC3  & CB1, CB2, CB3, CI1, CI2, CI3, CN\phantom{, CO1, CO2, CO3} & CO1, CO2, CO3 \\
\midrule
    \multirow{9}{*}{\begin{tabular}[c]{@{}c@{}}\textbf{Fault}\\ \textbf{severity}\end{tabular}}
    & TC4  & \phantom{CB1, }CB2, CB3, CI1, CI2, CI3, CN, CO1, CO2, CO3 & CB1 \\
    & TC5  & CB1, \phantom{CB1, }CB3, CI1, CI2, CI3, CN, CO1, CO2, CO3 & CB2 \\
    & TC6  & CB1, CB2, \phantom{CB1, }CI1, CI2, CI3, CN, CO1, CO2, CO3 & CB3 \\
    & TC7  & CB1, CB2, CB3, \phantom{CB1, }CI2, CI3, CN, CO1, CO2, CO3 & CI1 \\
    & TC8  & CB1, CB2, CB3, CI1, \phantom{CB1, }CI3, CN, CO1, CO2, CO3 & CI2 \\
    & TC9  & CB1, CB2, CB3, CI1, CI2, \phantom{CB1, }CN, CO1, CO2, CO3 & CI3 \\
    & TC10 & CB1, CB2, CB3, CI1, CI2, CI3, CN, \phantom{CB1, }CO2, CO3 & CO1 \\
    & TC11 & CB1, CB2, CB3, CI1, CI2, CI3, CN, CO1, \phantom{CB1, }CO3 & CO2 \\
    & TC12 & CB1, CB2, CB3, CI1, CI2, CI3, CN, CO1, CO2\phantom{CB1, } & CO3 \\
\bottomrule
\end{tabular}
\end{table}

Table~\ref{tab:Task C} summarizes our OSFD tasks. In each task, we designate either one fault type or fault severity class as the unknown class and treat the remaining classes as known classes, thereby evaluating the model's sensitivity to unanticipated fault signatures. This fine-grained protocol assigns each fault severity class as the unknown category in turn, enabling a granular assessment of whether the method can detect different unseen severity levels while maintaining known-class classification performance.

For reliability evaluations, we partition the dataset into training, validation, selection, and test sets. After applying the preprocessing pipeline described in Section~\ref{sec:MGDC}, the raw vibration sequences used for training, validation, selection, and testing contain 288,000, 48,000, 48,000, and 96,001 data points, respectively. The STFT-based preprocessing pipeline then transforms these raw signals into discrete high-dimensional feature instances. 


\subsubsection{STFT configuration selection of the CWRU dataset} \label{sec:STFT_C}

\begin{table}[ht!]
\caption{STFT configuration search space for preprocessing the CWRU dataset in fault severity diagnosis. The search space includes 38 candidate configurations obtained by varying the frequency-bin and time-step resolutions across signal lengths from 1 to 5 revolutions.} 
\label{tab:Search_space_C} 
\centering
\begin{tabular}{l cccccccc}
\toprule
    \textbf{Metric} & \multicolumn{8}{c}{\textbf{STFT configurations}} \\
\midrule
    \multicolumn{9}{c}{\textit{Signal length: 1 revolution}} \\
        Frequency bins    & -- & -- & 512 & 256 & 128 & 64 & 32 & 16 \\
    Time steps        & -- & -- & 4 & 10 & 23 & 50 & 100 & 200 \\
    Silhouette score  & -- & -- & 0.1228 & 0.1020 & 0.0871 & 0.0808 & 0.0617 & 0.0380 \\
    H-score           & -- & -- & 0.7211 & 0.7022 & 0.7603 & 0.4522 & 0.4350 & 0.3388 \\
\midrule
    \multicolumn{9}{c}{\textit{Signal length: 2 revolutions}} \\
        Frequency bins    & 2048 & 1024 & 512 & 256 & 128 & 64 & 32 & 16 \\
    Time steps        & 1 & 4 & 10 & 23 & 49 & 100 & 203 & 410 \\
    Silhouette score  & 0.1591 & 0.1404 & 0.1244 & 0.1052 & 0.0898 & 0.0830 & 0.0660 & 0.0392 \\
    H-score           & 0.8809 & 0.7854 & 0.9462 & 0.8601 & 0.6596 & 0.7117 & 0.5094 & 0.6246 \\
\midrule
    \multicolumn{9}{c}{\textit{Signal length: 3 revolutions}} \\
    Frequency bins    & 2048 & 1024 & 512 & 256 & 128 & 64 & 32 & 16 \\
    Time steps        & 2 & 7 & 16 & 36 & 74 & 152 & 307 & 616 \\
    Silhouette score  & 0.1593 & 0.1368 & 0.1277 & 0.1074 & 0.0912 & 0.0845 & 0.0672 & 0.0407 \\
    H-score           & 0.8615 & 0.9844 & 0.9948 & 0.6943 & 0.6033 & 0.6662 & 0.7446 & 0.5593 \\
\midrule
    \multicolumn{9}{c}{\textit{Signal length: 4 revolutions}} \\
    Frequency bins    & 2048 & 1024 & 512 & 256 & 128 & 64 & 32 & 16 \\
    Time steps        & 3 & 10 & 22 & 47 & 97 & 200 & 400 & 800 \\
    Silhouette score  & 0.1603 & 0.1369 & 0.1275 & 0.1077 & 0.0914 & 0.0845 & 0.0674 & 0.0411 \\
    H-score           & 0.9323 & 0.9781 & 0.8670 & 0.8636 & 0.6992 & 0.5670 & 0.8676 & 0.8398 \\
\midrule
    \multicolumn{9}{c}{\textit{Signal length: 5 revolutions}} \\
    Frequency bins    & 2048 & 1024 & 512 & 256 & 128 & 64 & 32 & 16 \\
    Time steps        & 5 & 13 & 30 & 60 & 125 & 250 & 500 & 1000 \\
    Silhouette score  & 0.1587 & 0.1352 & 0.1265 & 0.1075 & 0.0912 & 0.0844 & 0.0668 & 0.0400 \\
    H-score           & 0.9535 & 0.9729 & 0.9607 & 0.7126 & 0.7067 & 0.8063 & 0.6888 & 0.6988 \\
\bottomrule
\end{tabular}
\end{table}

To evaluate the effectiveness of the proposed MGDC preprocessing strategy, we construct a predefined STFT configuration search space for the CWRU dataset, as summarized in Table~\ref{tab:Search_space_C}. The search space includes 38 candidate STFT configurations generated by varying the signal length from 1 to 5 rotor revolutions and combining different frequency-bin and time-step resolutions. This design covers a range of time--frequency resolution conditions induced by the inherent trade-off in the STFT.

For each candidate STFT configuration, we first transform the raw vibration signals into two-dimensional spectrograms. We then apply the square-root scaling and frequency-bin-wise normalization described in Section~\ref{sec:MGDC} to obtain normalized spectrogram representations. For metric computation only, we flatten these representations into one-dimensional feature vectors and use the known fault severity labels as cluster assignments. Based on these feature representations, we evaluate separability metrics, including the Silhouette score, to rank the candidate STFT configurations before network training. Appendix~\ref{apdx:Metrics} provides the detailed definitions of the separability metrics.

\begin{figure}[ht!]
\centering 
\includegraphics[width=0.99\linewidth]{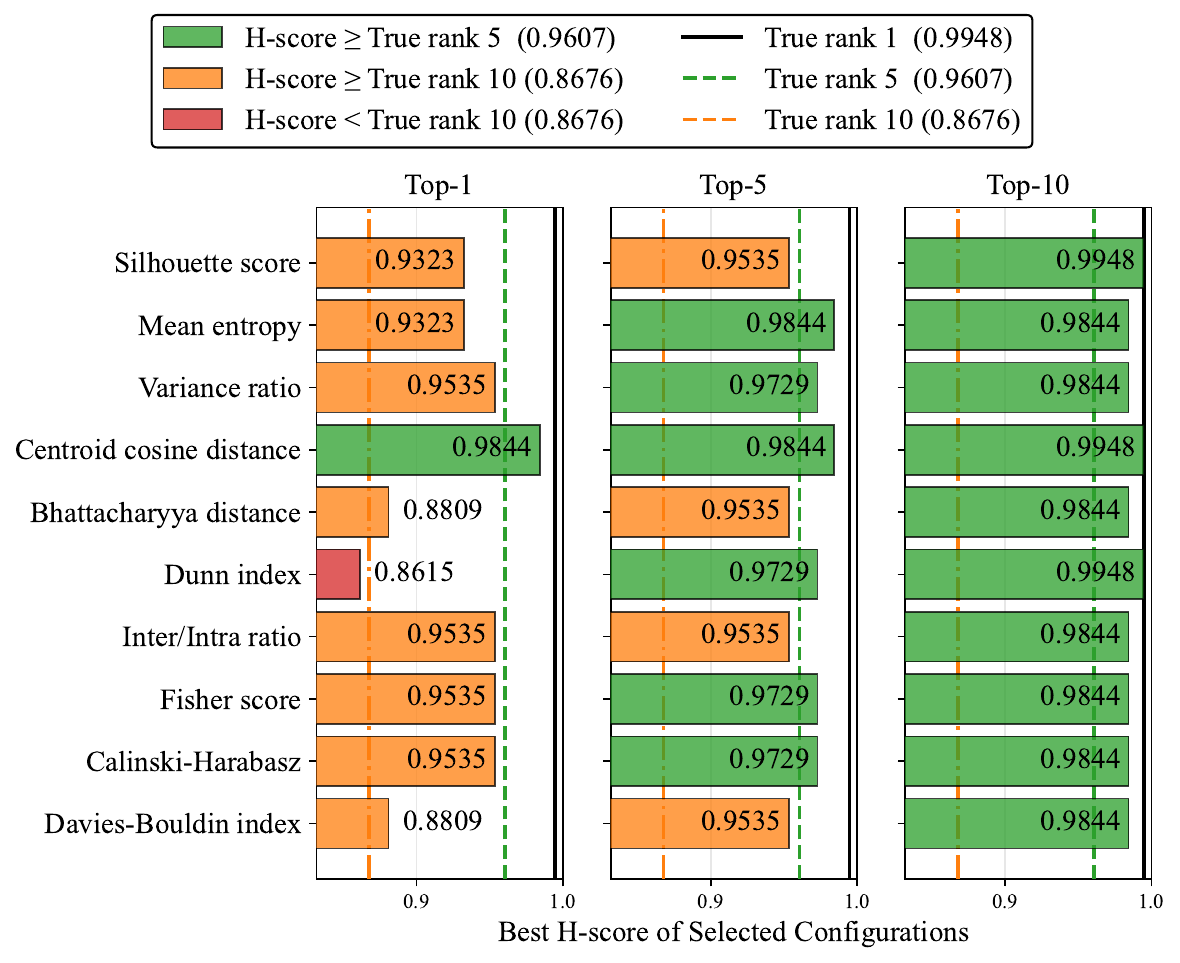} 
\caption{Top-$R$ H-score analysis for metric-guided STFT configuration selection on the CWRU dataset. Each bar shows the best H-score achieved within the reduced candidate set $\mathcal{H}_R$, and the vertical reference lines indicate the true H-score of the rank-1, rank-5, and rank-10 configurations. Green and orange bars denote cases where the best H-score exceeds the true rank-5 and rank-10, respectively.}
\label{fig:CWRU_avg_acc} 
\end{figure}

Fig.~\ref{fig:CWRU_avg_acc} compares the diagnostic performance of different separability metrics as the reduced candidate set $R$ increases. The best H-score is computed on the selection set. The results show that relying on only the top-1 configuration selected by a metric can be unstable, because the highest-ranked configuration in the metric space does not necessarily yield the best diagnostic performance. In contrast, retaining a small reduced candidate set allows several metrics to include high-performing STFT configurations among the top-ranked candidates. In particular, the Silhouette score, Centroid cosine distance, and Dunn index reach converged diagnostic performance at $R=10$, indicating that a compact candidate set can preserve effective STFT representations for subsequent network-based validation.

\begin{figure}[ht!]
\centering 
\includegraphics[width=0.6\linewidth]{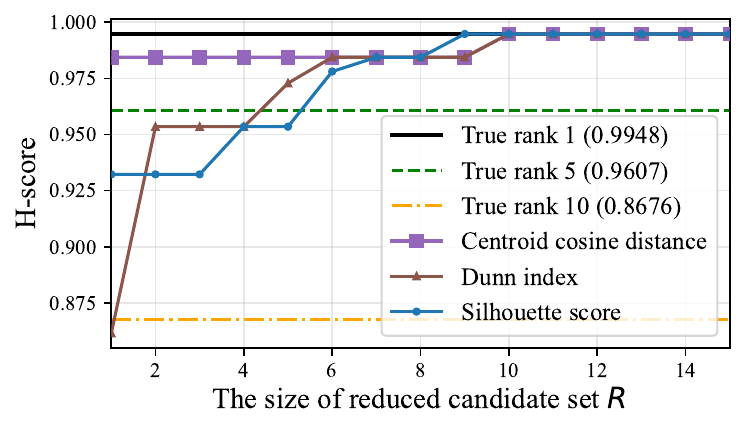} 
\caption{Convergence of the cumulative best H-score with respect to the size of the reduced candidate set $R$ on the CWRU dataset. The curves correspond to the three separability metrics: Centroid cosine distance, Dunn index, and Silhouette score. The horizontal lines denote the true H-scores of rank-1, rank-5, and rank-10. The Silhouette score reaches the true rank-1 H-score at $R=9$, whereas centroid cosine distance and Dunn index reach it at $R=10$.}
\label{fig:CWRU_convergence}
\end{figure}

\begin{figure}[ht!]
\centering 
\includegraphics[width=0.99\linewidth]{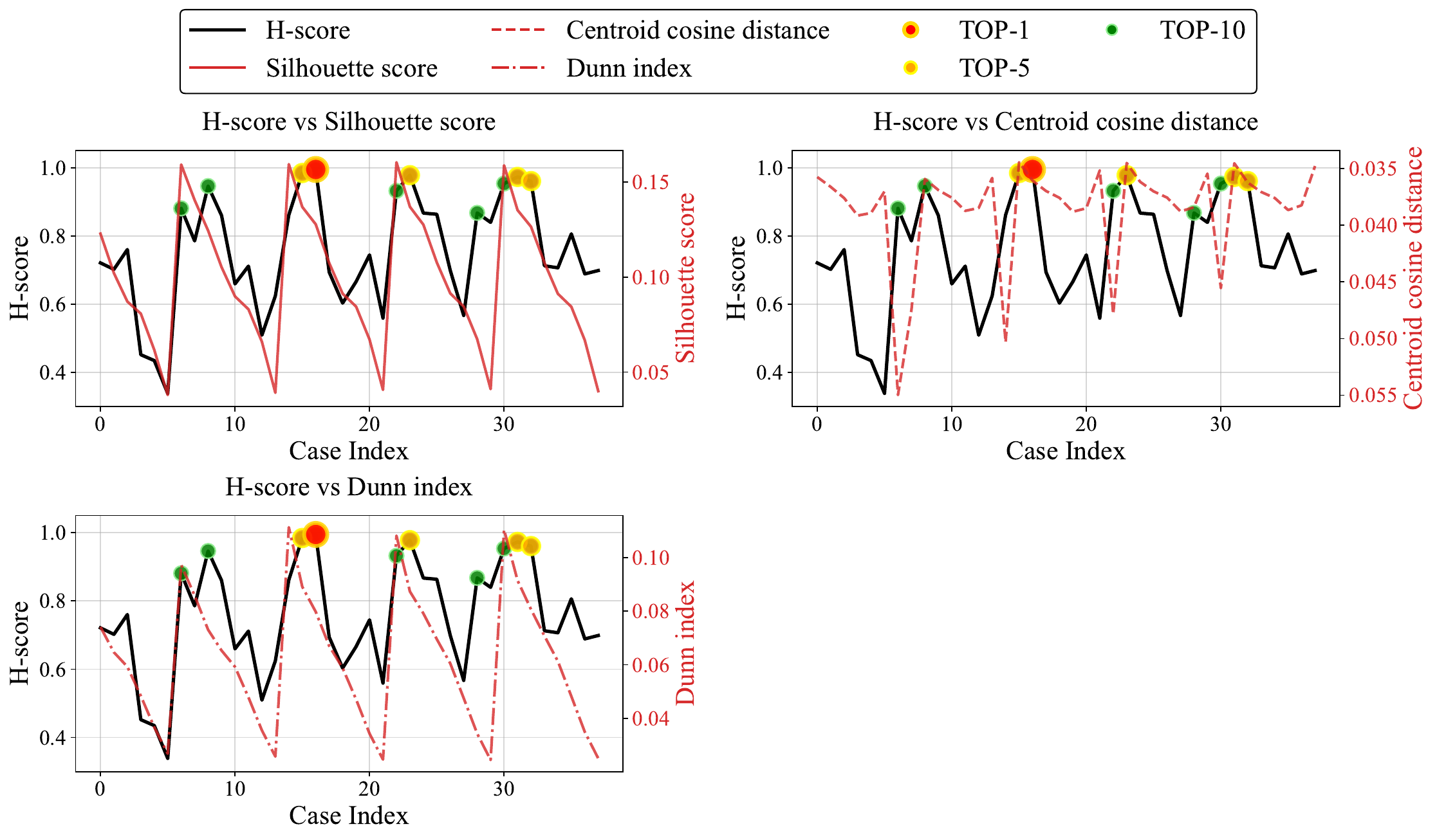}
\caption{Comparison between diagnostic performance and the separability metric across the STFT configuration search space $\mathcal{H}$ of the CWRU dataset. The blue line denotes the true H-score, while the red line represents the corresponding separability metric value of the configuration. The red, yellow, and green markers denote the true top-1, top-5, and top-10 candidate STFT configurations, respectively.}
\label{fig:CWRU_comparison} 
\end{figure}

Fig.~\ref{fig:CWRU_convergence} shows the cumulative best H-score as the reduced candidate-set size $R$ increases, with each candidate evaluated on the selection set. The Silhouette score identifies the globally best H-score of $0.9948$ at $R=9$, whereas the Centroid Cosine Distance and Dunn index require $R=10$. Fig.~\ref{fig:CWRU_comparison} compares the H-scores of the metric-selected candidates with their evaluated H-scores. 
 
\begin{table}[ht!]
\centering 
\caption{Computation time for various metrics of the CWRU dataset. The Silhouette score is selected as the primary metric by considering both early convergence behavior and computational cost.} 
\label{tab:Comp_time_C}
\begin{tabular}{lc @{\hspace{30pt}} lc}
\toprule
    \textbf{Metric} & \textbf{Time (s)} & \textbf{Metric} & \textbf{Time (s)} \\
\midrule
    Dunn Index               & 47.236 & Calinski-Harabasz        & 2.572 \\
    Variance ratio           & 6.220  & Davies-Bouldin Index     & 2.320 \\
    Inter/intra ratio        & 3.760  & Centroid cosine distance & 1.591 \\
    Mean entropy             & 3.441  & Silhouette score         & 1.333 \\
    Bhattacharyya Distance   & 2.850  & Fisher score             & 0.864 \\
\bottomrule
\end{tabular}
\end{table}

Table~\ref{tab:Comp_time_C} summarizes the computational cost of each separability metric. The Silhouette score and Centroid Cosine Distance require only $1.333$ and $1.591$~s, respectively. In contrast, the Dunn index requires $47.236$~s, making it $35.4$ times slower than the Silhouette score and $29.7$ times slower than the Centroid Cosine Distance. Based on both candidate-selection efficiency and computational cost, we use the Silhouette score as the ranking criterion in the proposed MGDC strategy. For the CWRU dataset, MGDC identifies the best-performing STFT configuration by validating only the nine highest-ranked candidates, reducing the number of network-based evaluations from 38 to 9. The selected configuration uses 512 frequency bins and 16 time steps and achieves the highest H-score.


\subsubsection{Diagnostic results of the CWRU dataset} \label{sec:Result_C}

\begin{table}[ht!] 
\centering 
\caption{Diagnostic performance comparison on the CWRU dataset under two diagnostic settings. CSAE achieves the best overall performance.} 
\label{tab:Result_C}
\begin{tabular}{c l ccccc}
\toprule
    \textbf{Granularity} & \textbf{Metric} & \textbf{Proposed} & \textbf{Global AE} & \textbf{OpenMax} & \textbf{CPL} & \textbf{ARPL} \\
\midrule
    \multirow{6}{*}{\shortstack{Fault \\ type}} 
    & micro F1 & 0.9416 & 0.8292 & 0.6431 & 0.8969 & 0.4366 \\
    & macro F1 & 0.9464 & 0.7721 & 0.5856 & 0.9121 & 0.3291 \\
    & UDA      & 0.8242 & 0.5447 & 0.2790 & 0.7684 & 0.2454 \\
    & CSA      & 0.9920 & 0.9511 & 0.7992 & 0.9520 & 0.5185 \\
    & OSA      & 0.9416 & 0.8292 & 0.6431 & 0.8969 & 0.4366 \\
    & H-score  & 0.8892 & 0.6822 & 0.3668 & 0.8454 & 0.2905 \\
\midrule
    \multirow{6}{*}{\shortstack{Fault \\ severity}} 
    & micro F1 & 0.9902 & 0.7552 & 0.1941 & 0.8703 & 0.3190 \\
    & macro F1 & 0.9904 & 0.7275 & 0.1204 & 0.8641 & 0.2244 \\
    & UDA      & 0.9953 & 0.9060 & 0.1093 & 0.3682 & 0.2270 \\
    & CSA      & 0.9896 & 0.7384 & 0.2035 & 0.9261 & 0.3293 \\
    & OSA      & 0.9902 & 0.7552 & 0.1941 & 0.8703 & 0.3190 \\
    & H-score  & 0.9924 & 0.8015 & 0.0826 & 0.4449 & 0.1659 \\
\bottomrule
\end{tabular}
\end{table}

In this section, we compare the OSFD performance of the proposed method on the test set with the baseline methods presented in Section~\ref{sec:Baseline}. To isolate the effect of the diagnostic architecture, we evaluate all baseline methods using the same MGDC-selected STFT representation as the proposed method. Table~\ref{tab:Result_C} reports the results for two diagnostic granularities: coarse-grained fault type diagnosis and fine-grained fault severity diagnosis.

In the fault type diagnosis, the proposed method achieves the best overall performance across all evaluation metrics. Specifically, it obtains the highest micro F1, macro F1, UDA, CSA, OSA, and H-score values of 0.9416, 0.9464, 0.8242, 0.9920, 0.9416, and 0.8892, respectively. Compared with CPL, which achieves the second-highest H-score of 0.8454, the proposed method improves the H-score by 0.0438. This result indicates that the class-specific reconstruction structure of the proposed CSAE-based architecture is also effective for coarse-grained fault type diagnosis. In particular, the proposed method maintains a favorable balance between known-class classification accuracy and unknown-class detection, whereas the baseline methods show either reduced UDA or degraded CSA.

The advantage of the proposed method becomes more pronounced in the fault severity diagnosis, which requires the model to separate highly overlapping latent manifolds formed by adjacent degradation states. Baseline models that rely on a single shared representation or global decision boundary struggle to maintain balanced performance between CSA and UDA. For example, OpenMax and ARPL show severe degradation in unknown-class detection, yielding UDA values of 0.1093 and 0.2270, respectively. CPL achieves a relatively high CSA of 0.9261 but suffers from a low UDA of 0.3682, resulting in an H-score of 0.4449. Global AE achieves a higher UDA of 0.9060 than the other baselines, but its CSA decreases to 0.7384, leading to an H-score of 0.8015. In contrast, the proposed method achieves both high UDA and CSA values of 0.9953 and 0.9896, respectively. As a result, it obtains the highest H-score of 0.9924 in the fine-grained fault severity diagnosis. These results demonstrate that the proposed CSAE-based architecture effectively mitigates feature entanglement among adjacent severity states and provides a balanced OSFD solution across both coarse-grained and fine-grained diagnostic settings.

\begin{figure}[ht!]
\centering
\includegraphics[width=\textwidth]{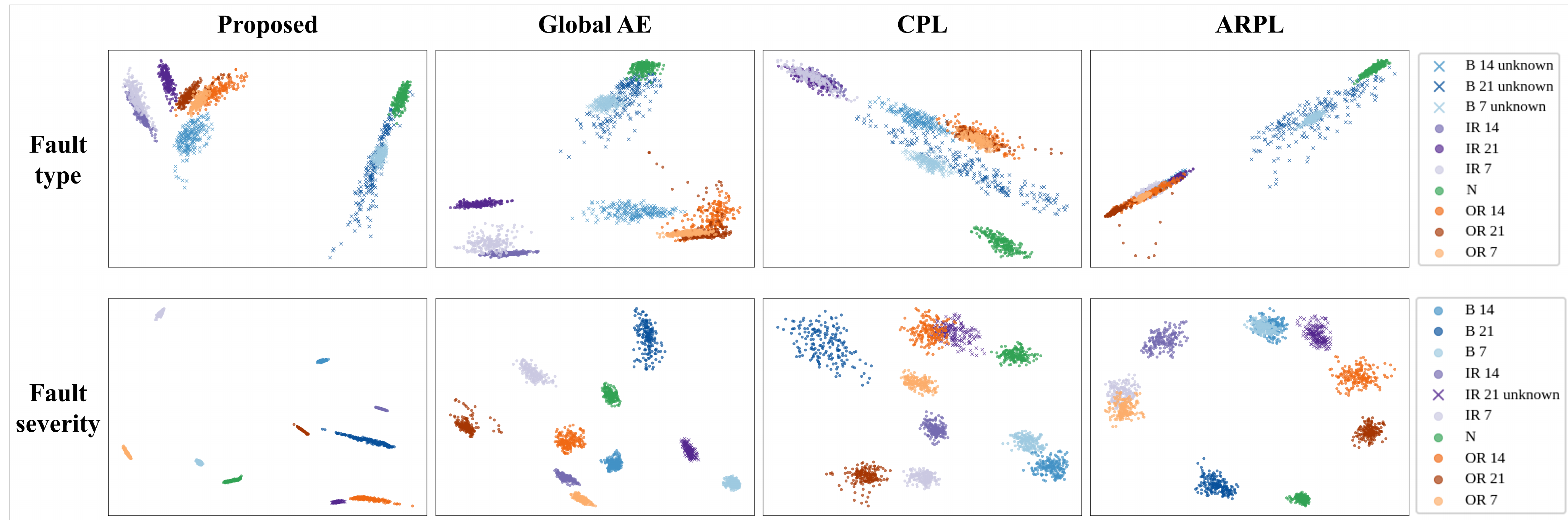}
\caption{t-distributed stochastic neighbor embedding (t-SNE) visualization of the CWRU dataset. In fault type diagnosis setting, B7, B14, and B21 are treated as unknown classes. Samples sharing the same fault type but having different severity levels tend to form type-level groups. In fault severity diagnosis setting, IR21 is treated as unknown class. The proposed method produces more compact and clearly separated severity-specific clusters than the baseline methods.}
\label{fig:tSNE_C}
\end{figure}

Fig.~\ref{fig:tSNE_C} shows the feature distributions obtained using t-distributed stochastic neighbor embedding (t-SNE). In fault type diagnosis, severity levels associated with the same fault type are grouped into a single class, producing broad manifolds that span multiple degradation stages. Although these representations distinguish fault types, they do not explicitly separate severity variations within the same fault mechanism. In fault-severity diagnosis, each severity level is treated as an independent class. The proposed method produces compact, well-separated severity-specific clusters, whereas the baseline methods yield more dispersed or entangled distributions. This visualization suggests that the class-specific reconstruction structure of the CSAE reduces feature entanglement and improves fine-grained class separability.

\begin{figure}[ht!]
\centering
\includegraphics[width=0.95\textwidth]{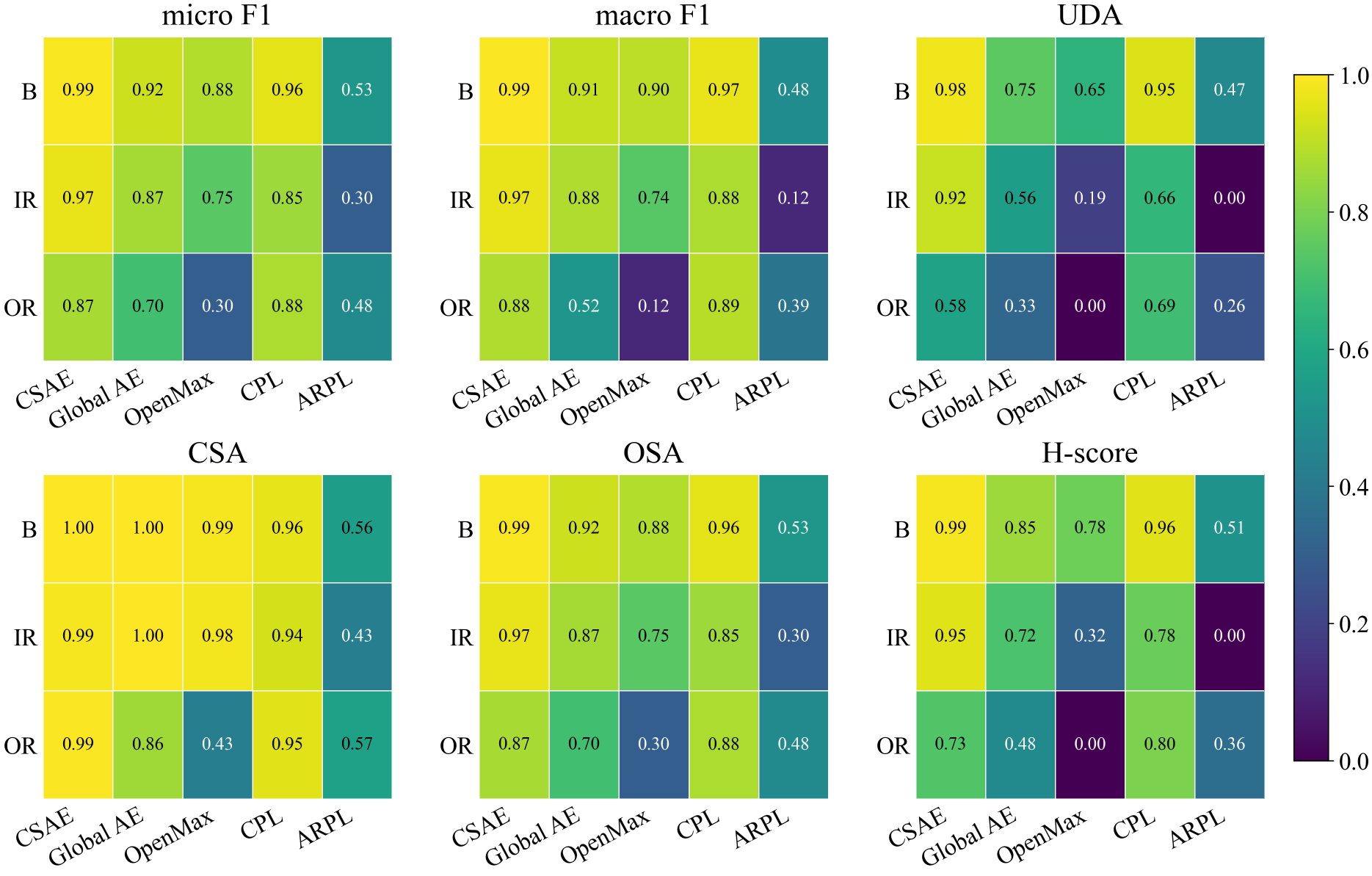}
\caption{Task-wise performance heatmaps for fault type diagnosis on the CWRU dataset. Rows indicate the unknown fault type, and columns indicate the compared methods.}
\label{fig:score_CWRU_type}
\end{figure}

\begin{figure}[ht!]
\centering
\includegraphics[width=0.95\textwidth]{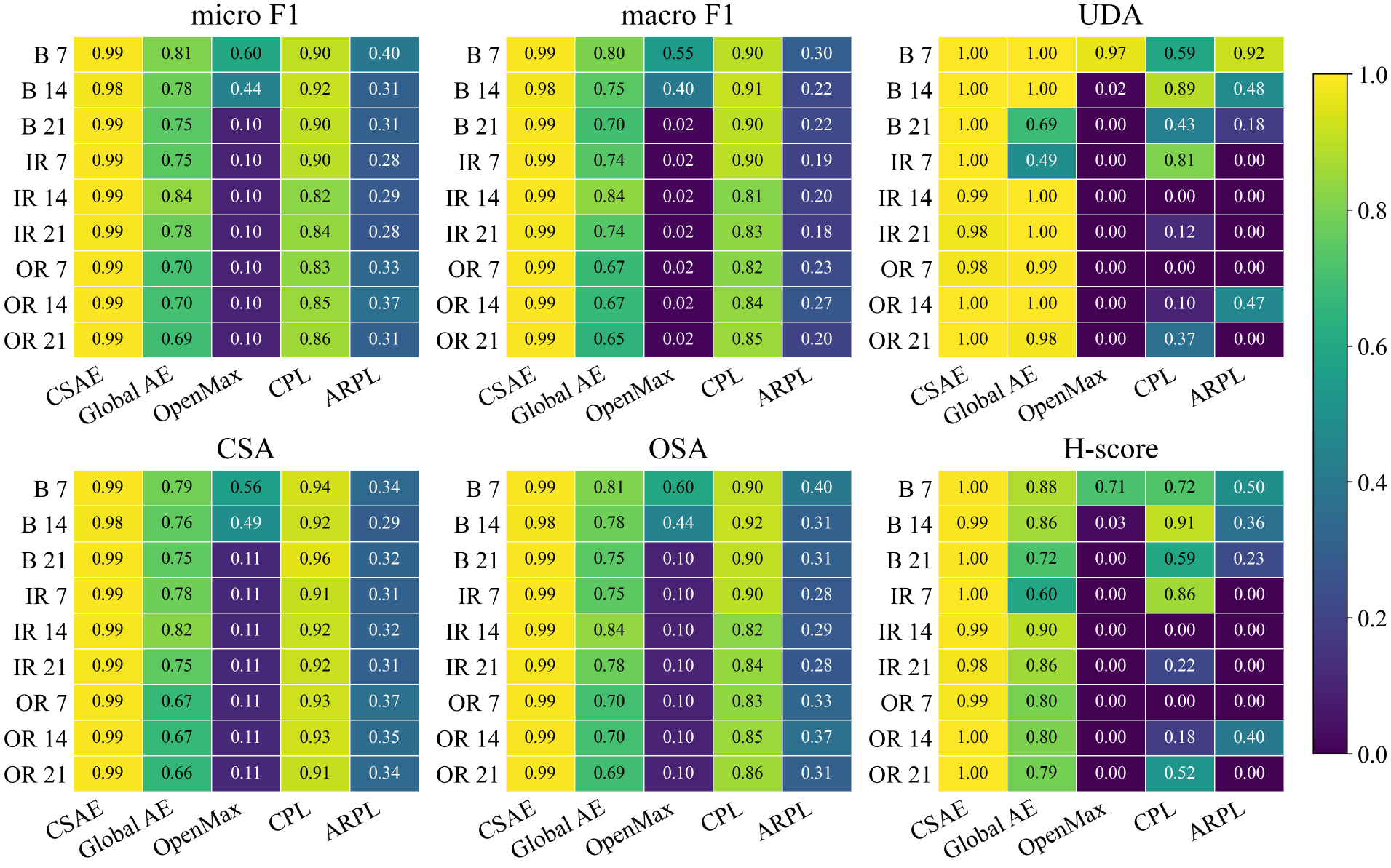}
\caption{Task-wise performance heatmaps for fault severity diagnosis on the CWRU dataset. Rows indicate the unknown fault severity, and columns indicate the compared methods.}
\label{fig:score_CWRU_severity}
\end{figure}

Figs~\ref{fig:score_CWRU_type} and \ref{fig:score_CWRU_severity} provide a more detailed comparison of task-specific diagnostic performance. In fault type diagnosis, the proposed method maintains high scores across ball (B), inner-race (IR), and outer-race (OR) faults, with only a slight decrease for the OR fault. In contrast, the baseline methods exhibit greater performance variation across the unknown fault types.

The proposed method achieves consistently high UDA, CSA, OSA, and H-scores across nearly all severity states. In contrast, the baseline methods exhibit severity-dependent degradation; for example, OpenMax and ARPL show relatively poor unknown detection, while CPL and the Global AE show larger variations across severity states. These results indicate that the proposed method improves both average diagnostic performance and consistency in fine-grained fault severity diagnosis.


\subsection{Case 2: PU dataset} \label{sec:PU}

This case study evaluates the proposed method on the Paderborn University (PU) bearing dataset~\citep{lessmeier2016condition} to verify its applicability to a bearing system different from the CWRU dataset. The objective is to examine whether the proposed method consistently maintains known-class classification performance and rejects unseen fault states under both fault type-level and fault severity-level OSFD scenarios.

\subsubsection{Dataset and task description of the PU dataset} \label{sec:Dataset P}

\begin{table}[ht!]
\centering 
\caption{Specifications of the PU dataset.} 
\label{tab:Dataset P}
\renewcommand{\arraystretch}{1.2}
\begin{tabular}{cccc}
\toprule
    \textbf{Class} & \textbf{Fault type} & \textbf{Fault severity} & \textbf{Bearing code} \\ 
\midrule
    PN  & Normal                    & -   & K004 \\ 
    PA1 & Outer race fault          & 1 & KA04 \\ 
    PA2 & Outer race fault          & 2 & KA16 \\ 
    PB1 & Inner \& outer race fault & 1 & KB27 \\ 
    PB2 & Inner \& outer race fault & 2 & KB23 \\ 
    PI1 & Inner race fault          & 1 & KI21 \\ 
    PI2 & Inner race fault          & 2 & KI18 \\
\bottomrule
\end{tabular}
\end{table}

\begin{table}[ht!]
\centering 
\caption{Configuration of OSFD tasks for the PU dataset. Tasks TP1--TP3 focus on unknown fault types, while TP4--TP9 address the diagnosis of unseen fault severity levels.} 
\label{tab:Task P}
\renewcommand{\arraystretch}{1.2}
\begin{tabular}{cccc}
\toprule
    \textbf{Granularity} & \textbf{Task} & \textbf{Known class} & \textbf{Unknown class} \\ 
\midrule
    \multirow{3}{*}{\begin{tabular}[c]{@{}c@{}}\textbf{Fault}\\ \textbf{type}\end{tabular}} 
    & TP1 & PN, \phantom{PA1, PA2, }PB1, PB2, PI1, PI2 & PA1, PA2 \\
    & TP2 & PN, PA1, PA2, \phantom{PB1, PB2, }PI1, PI2 & PB1, PB2 \\
    & TP3 & PN, PA1, PA2, PB1, PB2\phantom{, PI1, PI2} & PI1, PI2 \\
\midrule
    \multirow{6}{*}{\begin{tabular}[c]{@{}c@{}}\textbf{Fault}\\ \textbf{severity}\end{tabular}} 
    & TP4 & PN, \phantom{PA1, }PA2, PB1, PB2, PI1, PI2 & PA1 \\
    & TP5 & PN, PA1, \phantom{PA2, }PB1, PB2, PI1, PI2 & PA2 \\
    & TP6 & PN, PA1, PA2, \phantom{PB1, }PB2, PI1, PI2 & PB1 \\
    & TP7 & PN, PA1, PA2, PB1, \phantom{PB2, }PI1, PI2 & PB2 \\
    & TP8 & PN, PA1, PA2, PB1, PB2, \phantom{PI1, }PI2 & PI1 \\
    & TP9 & PN, PA1, PA2, PB1, PB2, PI1\phantom{, PI2} & PI2 \\
\bottomrule
\end{tabular} 
\end{table}

The PU dataset contains realistic bearing damages that form a continuous and dense fault manifold, which can blur the decision boundaries between known and unknown classes. This characteristic makes the PU dataset suitable for evaluating OSFD performance under more realistic fault conditions. Consistent with the CWRU setup, we apply a fine-grained evaluation protocol in which either one fault type or one fault severity is designated as the unknown class while the remaining conditions are treated as known classes, as summarized in Table~\ref{tab:Task P}. This protocol enables a granular assessment of the model's diagnostic reliability under realistic open-set fault conditions. Table~\ref{tab:Dataset P} provides the detailed dataset description. In contrast to the CWRU dataset, where severity is defined by defect diameter in inches, the PU dataset describes realistic bearing damages using ordinal damage levels without a fixed physical unit.

For the experimental implementation, we partition the raw vibration sequences into 3,071,964 training, 511,994 validation, 511,993 selection, and 1,023,989 test data points. After applying the preprocessing pipeline, these sequences are transformed into high-dimensional feature representations for deep learning. 


\subsubsection{STFT configuration selection of the PU dataset} \label{sec:STFT_P}

\begin{table}[ht!]
\caption{STFT configuration search space for preprocessing the PU dataset. The search space includes 39 candidate configurations obtained by varying the frequency-bin and time-step resolutions across signal lengths from 1 to 5 revolutions.}
\label{tab:Search_space_P}  
\centering
\begin{tabular}{l cccccccc}
\toprule
    \textbf{Metric} & \multicolumn{8}{c}{\textbf{STFT configurations}} \\
\midrule
    \multicolumn{9}{c}{\textit{Signal length: 1 revolution}} \\
    Frequency bins   & -- & 1024 & 512 & 256 & 128 & 64 & 32 & 16 \\
    Time steps       & -- & 2 & 7 & 17 & 37 & 77 & 157 & 317 \\
    Silhouette score & -- & 0.0546 & 0.0512 & 0.0421 & 0.0357 & 0.0296 & 0.0201 & 0.0122 \\
    H-score          & -- & 0.5939 & 0.6161 & 0.6130 & 0.4743 & 0.4157 & 0.6867 & 0.2761 \\
\midrule
    \multicolumn{9}{c}{\textit{Signal length: 2 revolutions}} \\
    Frequency bins   & 2048 & 1024 & 512 & 256 & 128 & 64 & 32 & 16 \\
    Time steps       & 2 & 7 & 17 & 37 & 77 & 157 & 317 & 637 \\
    Silhouette score & 0.0590 & 0.0579 & 0.0519 & 0.0427 & 0.0362 & 0.0296 & 0.0202 & 0.0123 \\
    H-score          & 0.8563 & 0.9005 & 0.9790 & 0.8946 & 0.8575 & 0.7395 & 0.9121 & 0.6852 \\
\midrule
    \multicolumn{9}{c}{\textit{Signal length: 3 revolutions}} \\
    Frequency bins   & 2048 & 1024 & 512 & 256 & 128 & 64 & 32 & 16 \\
    Time steps       & 5 & 12 & 27 & 57 & 117 & 237 & 477 & 957 \\
    Silhouette score & 0.0595 & 0.0585 & 0.0522 & 0.0430 & 0.0364 & 0.0297 & 0.0203 & 0.0124 \\
    H-score          & 0.9772 & 0.9429 & 0.9615 & 0.9035 & 0.9882 & 0.7805 & 0.7953 & 0.7496 \\
\midrule
    \multicolumn{9}{c}{\textit{Signal length: 4 revolutions}} \\
    Frequency bins   & 2048 & 1024 & 512 & 256 & 128 & 64 & 32 & 16 \\
    Time steps       & 7 & 17 & 37 & 77 & 157 & 317 & 637 & 1277 \\
    Silhouette score & 0.0596 & 0.0587 & 0.0524 & 0.0431 & 0.0365 & 0.0296 & 0.0203 & 0.0124 \\
    H-score          & 0.9883 & 0.7911 & 0.8692 & 0.7095 & 0.9549 & 0.6007 & 0.8057 & 0.8537 \\
\midrule
    \multicolumn{9}{c}{\textit{Signal length: 5 revolutions}} \\
    Frequency bins   & 2048 & 1024 & 512 & 256 & 128 & 64 & 32 & 16 \\
    Time steps       & 10 & 22 & 47 & 97 & 197 & 397 & 797 & 1597 \\
    Silhouette score & 0.0596 & 0.0589 & 0.0523 & 0.0431 & 0.0365 & 0.0297 & 0.0204 & 0.0124 \\
    H-score          & 0.7861 & 0.9645 & 0.8522 & 0.8421 & 0.9829 & 0.8597 & 0.9871 & 0.9495 \\
\bottomrule
\end{tabular}
\end{table}

To further examine the consistency of the proposed MGDC preprocessing strategy, we apply the same STFT configuration selection procedure to the PU dataset. As summarized in Table~\ref{tab:Search_space_P}, the search space includes 39 candidate STFT configurations generated by varying the signal length from 1 to 5 revolutions and combining different frequency-bin and time-step resolutions. The number of candidate configurations differs from that of the CWRU dataset because the two datasets have different sampling frequencies and rotational speeds. These differences change the number of data points contained in 1 to 5 revolutions, resulting in a slightly different set of feasible frequency-bin and time-step combinations. Because the PU dataset contains realistic bearing damages with more complex and densely distributed feature manifolds, this case further evaluates the consistency of the Silhouette-score-based screening strategy under realistic bearing damage conditions.

For each candidate STFT configuration, we first convert the vibration signals into spectrograms and then apply the same square-root scaling and frequency-bin-wise normalization.  For metric computation only, we flatten the resulting normalized spectrogram representations into one-dimensional feature vectors and use the known fault severity labels as cluster assignments. Based on these representations, we evaluate several separability metrics to rank the candidate STFT configurations before network training.

\begin{figure}[ht!]
\centering 
\includegraphics[width=0.99\linewidth]{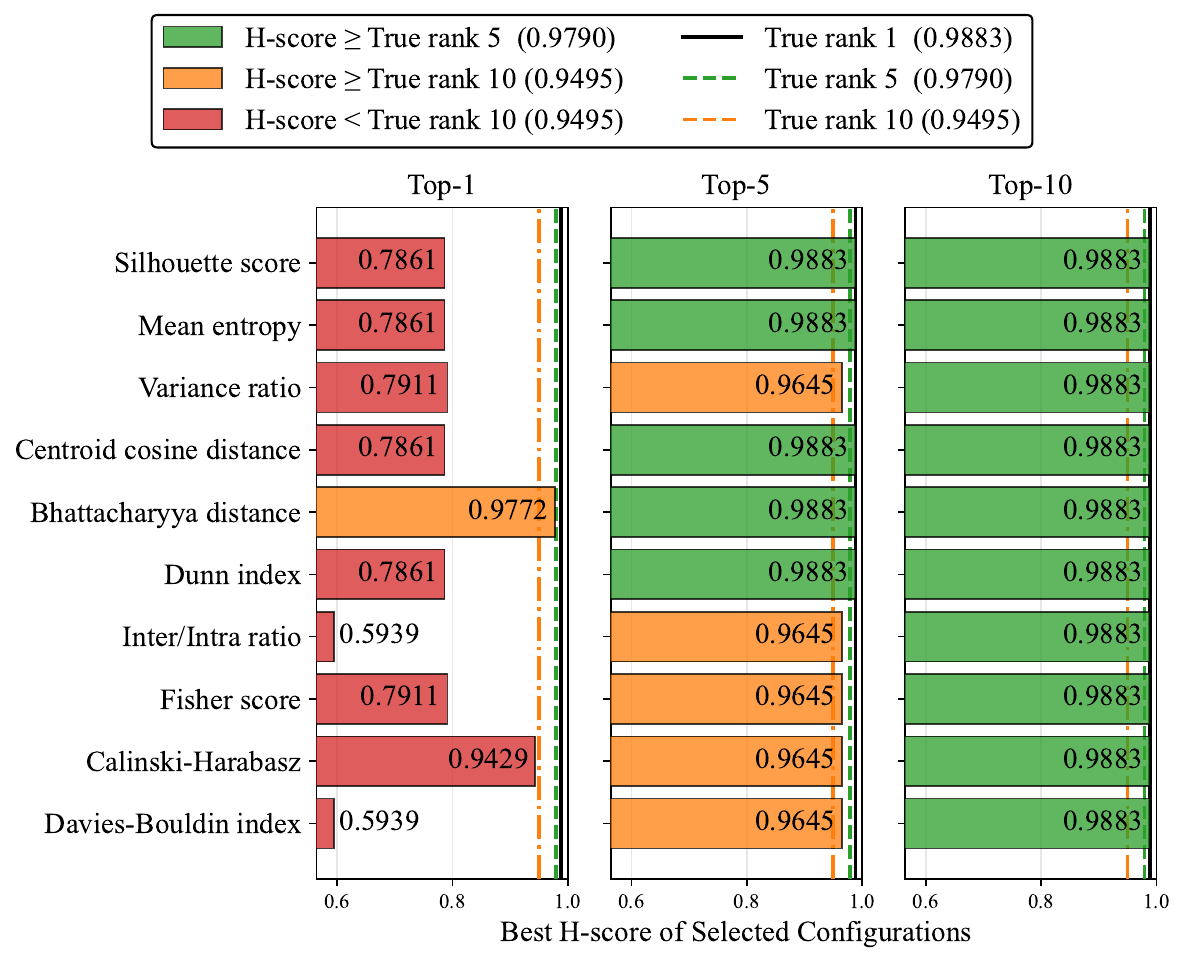}
\caption{Top-$R$ H-score analysis for metric-guided STFT configuration screening on the PU dataset. The horizontal bars denote the maximum H-score obtained from the reduced candidate set selected by each separability metric for a given reduced set size $R$. The black, green, and orange vertical lines indicate the globally best configuration, the top-5 global rank, and the top-10 global rank based on the H-score, respectively. Green bars indicate that the metric-selected reduced candidate set includes at least one configuration within the top-5 global rank, whereas orange bars indicate inclusion within the top-10 global rank. The results show that using a small reduced candidate set $\mathcal{H}_R$, rather than relying on a single top-1 configuration, improves the reliability of STFT configuration selection while substantially reducing the number of network-based validation trials.} 
\label{fig:PU_avg_acc} 
\end{figure}

Fig.~\ref{fig:PU_avg_acc} compares the screening performance of different separability metrics according to the size of the reduced candidate set $R$. The best H-score is computed on the selection set. The results show that several metrics can include high-performing STFT configurations even with a small reduced candidate set. In particular, the PU dataset exhibits rapid convergence from the early stage of the ranking process, indicating that diagnostically effective STFT representations can be identified without exhaustively validating all candidate configurations. This result further supports the use of a reduced candidate set for efficient STFT configuration selection.

\begin{figure}[ht!]
\centering 
\includegraphics[width=0.6\linewidth]{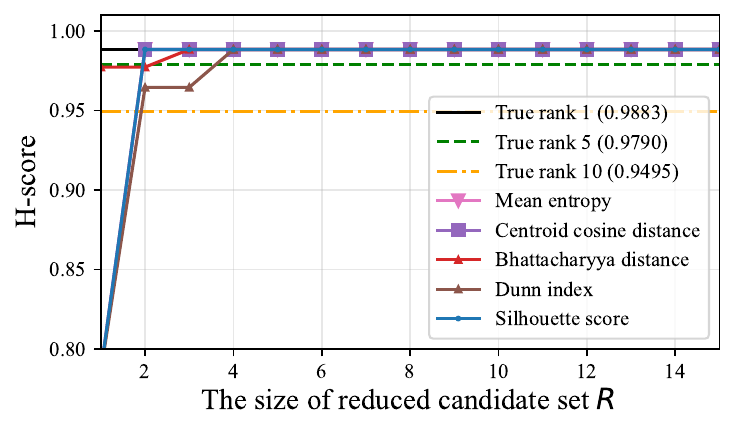}
\caption{Convergence of the cumulative best H-score with respect to the size of the reduced candidate set $R$ on the PU dataset. The curves correspond to the three separability metrics whose ranked candidate lists recover the globally best STFT configuration within the top-10 candidates: centroid cosine distance, Dunn index, and Silhouette score. The horizontal reference lines denote the H-scores of the true rank-1, rank-5, and rank-10 configurations obtained from the exhaustive search. The Silhouette score, Mean entropy, and Centroid cosine distance reach the true rank-1 H-score at $R=2$, Bhattacharyya distance reaches it at $R=3$, and Dunn index reaches it at $R=4$.}
\label{fig:PU_convergence} 
\end{figure}

\begin{figure}[ht!]
\centering 
\includegraphics[width=0.99\linewidth]{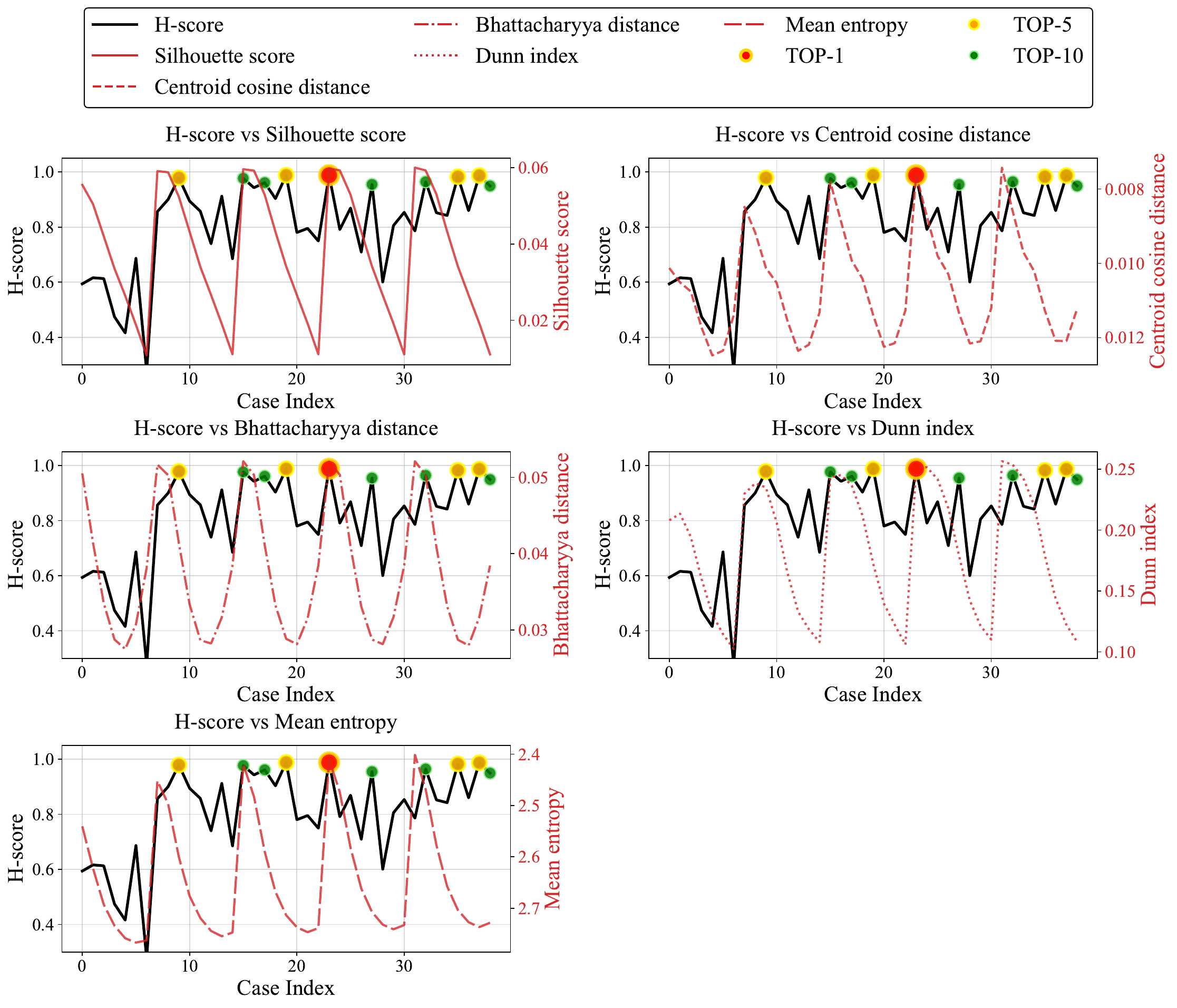}
\caption{Comparison between diagnostic performance and metric-based rankings across the STFT configuration search space $\mathcal{H}$ of the PU dataset. The blue solid line represents the H-score obtained from the diagnostic model, while the red line represents the corresponding separability metric value. The red, yellow, and green markers denote the top-1, top-5, and top-10 candidate STFT configurations selected by each metric, respectively. The results show that a small reduced candidate set $\mathcal{H}_R$ selected by the metric ranking can retain near-optimal STFT configurations while substantially reducing the number of network-training trials.}
\label{fig:PU_comparison} 
\end{figure}

\begin{table}[ht!]
\centering 
\caption{Computation time for various metrics of the PU dataset. The Silhouette score is selected as the primary metric by considering both early convergence behavior and computational cost.} 
\label{tab:Comp_time_P}
\begin{tabular}{lc @{\hspace{30pt}} lc}
\toprule
    \textbf{Metric} & \textbf{Time (s)} & \textbf{Metric} & \textbf{Time (s)} \\ 
\midrule
    Dunn Index               & 849.722 & Bhattacharyya Distance   & 34.181 \\
    Variance ratio           & 94.772  & Calinski-Harabasz        & 33.270 \\
    Inter/intra ratio        & 51.548  & Davies-Bouldin Index     & 26.707 \\
    Silhouette score         & 43.853  & Centroid cosine distance & 21.508 \\
    Mean entropy             & 34.849  & Fisher score             & 7.452 \\ 
\bottomrule
\end{tabular}
\end{table}

Fig.~\ref{fig:PU_convergence} shows the cumulative best H-score  as the reduced candidate-set size $R$ increases, with each candidate evaluated on the selection set. The Silhouette score, Mean entropy, and Centroid cosine distance identify the globally best H-score at $R=2$, whereas, Bhattacharyya distance reaches it at $R=3$, and the Dunn index require $R=4$. Fig.~\ref{fig:PU_comparison} compares the H-score with the true H-score. Table~\ref{tab:Comp_time_P} summarizes the computation time required by each separability metric, specifically the Silhouette score requires 43.853 s. Although some metrics, such as Fisher score and Centroid cosine distance, require lower computation time, the Silhouette score provides a favorable balance among early convergence behavior, ranking reliability, and computational efficiency.

Considering the results for both datasets, we use the Silhouette score as the primary ranking criterion in the proposed MGDC strategy. For the PU dataset, MGDC identifies the best-performing STFT configuration by validating only the two highest-ranked candidates, reducing the number of network-based evaluations from 39 to 2. The selected configuration uses 2048 frequency bins and seven time steps and achieves the highest H-score. These results demonstrate the consistent selection efficiency of Silhouette-based MGDC across the CWRU and PU datasets.


\subsubsection{Diagnostic results of the PU dataset} \label{sec:Result_P}

\begin{table}[ht!]
\centering 
\caption{Diagnostic performance comparison on the PU dataset under two diagnostic settings. In coarse-grained fault type diagnosis, the baseline methods remain competitive, with CPL achieving the highest H-score. In contrast, the proposed method achieves the best balanced performance in fine-grained fault severity diagnosis, particularly for the H-score.} 
\label{tab:Result_P}
\begin{tabular}{c l ccccc}
\toprule
    \textbf{Granularity} & \textbf{Metric} & \textbf{Proposed} & \textbf{Global AE} & \textbf{OpenMax} & \textbf{CPL} & \textbf{ARPL} \\
\midrule
    \multirow{6}{*}{\shortstack{Fault \\ type}} 
    & micro F1 & 0.9435 & 0.7553 & 0.2857 & 0.9431 & 0.7024 \\
    & macro F1 & 0.9470 & 0.6552 & 0.1111 & 0.9473 & 0.6612 \\
    & UDA      & 0.8065 & 0.4790 & 0.0000 & 0.9994 & 0.0937 \\
    & CSA      & 0.9982 & 0.8658 & 0.4000 & 0.9205 & 0.9459 \\
    & OSA      & 0.9435 & 0.7553 & 0.2857 & 0.9431 & 0.7024 \\
    & H-score  & 0.8698 & 0.5461 & 0.0000 & 0.9583 & 0.1690 \\
\midrule
    \multirow{6}{*}{\shortstack{Fault \\ severity}} 
    & micro F1 & 0.9846 & 0.9038 & 0.1429 & 0.9098 & 0.6462 \\
    & macro F1 & 0.9844 & 0.8860 & 0.0057 & 0.9151 & 0.5712 \\
    & UDA      & 0.9182 & 0.6909 & 0.0000 & 0.8837 & 0.1343 \\
    & CSA      & 0.9957 & 0.9392 & 0.1667 & 0.9142 & 0.7315 \\
    & OSA      & 0.9846 & 0.9038 & 0.1429 & 0.9098 & 0.6462 \\
    & H-score  & 0.9509 & 0.7624 & 0.0000 & 0.8822 & 0.1347 \\
\bottomrule
\end{tabular}
\end{table}

Table~\ref{tab:Result_P} summarizes the diagnostic performance of the PU dataset under two diagnostic settings on the test set. In fault type diagnosis, the proposed method achieves the highest micro F1-score, CSA, and OSA values of 0.9435, 0.9982, and 0.9435, respectively. However, CPL obtains the highest UDA and H-score values of 0.9994 and 0.9583, respectively. The proposed method achieves an H-score of 0.8698, which is 0.0885 lower than that of CPL. This result indicates that, for coarse-grained fault type diagnosis on the PU dataset, CPL provides a stronger unknown rejection boundary, whereas the proposed method provides higher known-class classification accuracy. Thus, the proposed method remains competitive for known-class recognition but does not achieve the best balanced open-set performance in this setting. 

In fine-grained fault severity diagnosis, the proposed method achieves the best overall performance across all evaluation metrics. Specifically, it obtains the highest micro F1-score, macro F1-score, UDA, CSA, OSA, and H-score values of 0.9846, 0.9844, 0.9182, 0.9957, 0.9846, and 0.9509, respectively. Compared with CPL, which shows the second-highest H-score of 0.8822, the proposed method improves the H-score by 0.0687. Although CPL achieves relatively high UDA and CSA values of 0.8837 and 0.9142, respectively,  its overall classification performance remains lower than that of the proposed method. Global AE also shows limited performance, with an H-score of 0.7624, while OpenMax and ARPL exhibit severe degradation, indicating their limitations in handling densely overlapping severity manifolds. 

The proposed method maintains both high unknown detection capability and high known-class classification accuracy, with a UDA of 0.9182 and a CSA of 0.9957. This balanced performance is attributed to the CSAE-based class-specific reconstruction structure and the dual-criteria anomaly rejection mechanism, which jointly reduce feature entanglement and improve unknown fault rejection. These results demonstrate that the proposed method provides a reliable solution for fine-grained open-set fault severity diagnosis under realistic bearing fault conditions.

\begin{figure}[ht!]
\centering
\includegraphics[width=\textwidth]{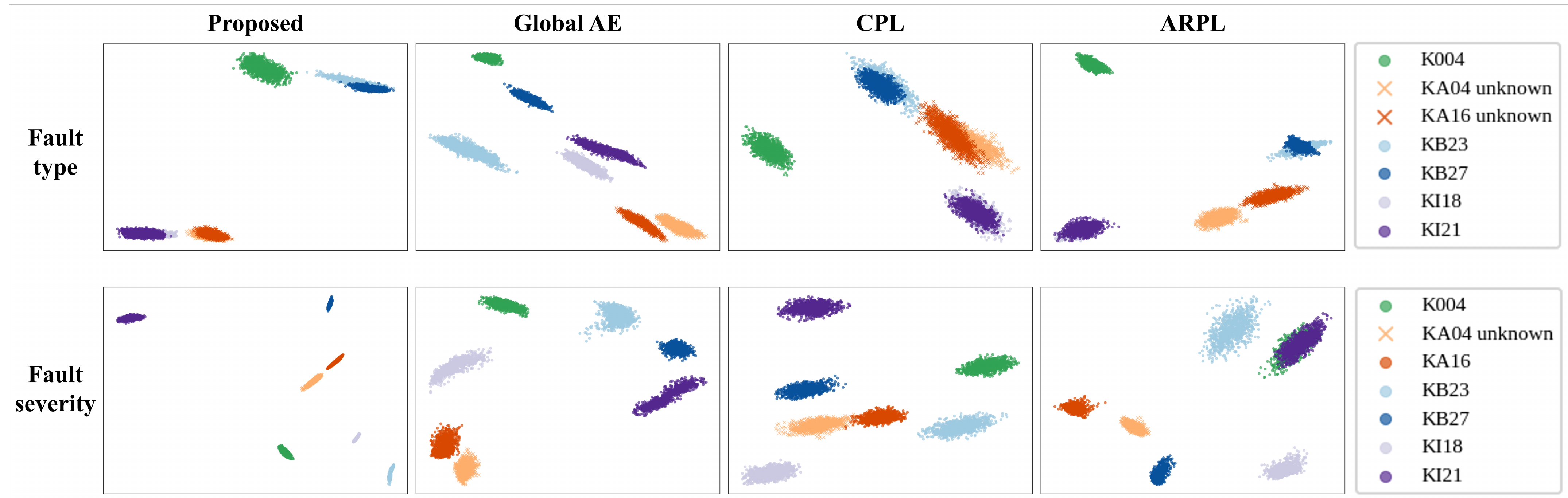}
\caption{t-distributed stochastic neighbor embedding (t-SNE) visualization of the PU dataset. In fault type diagnosis setting, KA04 and KA16 are treated as unknown classes. Samples sharing the same fault type but having different severity levels tend to form type-level groups. In fault severity diagnosis setting, KA04 is treated as unknown class. The proposed method produces more compact and clearly separated severity-specific clusters than the baseline methods.}
\label{fig:tSNE_P}
\end{figure}

Fig.~\ref{fig:tSNE_P} shows the t-SNE visualization of the learned feature distributions for the PU dataset. In fault-type diagnosis, different severity levels associated with the same fault mechanism are grouped into a single type-level class. In fault-severity diagnosis, the proposed method forms compact, well-separated severity-specific clusters, whereas the baseline methods produce more dispersed or partially entangled distributions. This result suggests that the CSAE architecture learns more discriminative class-specific manifolds under the realistic bearing-damage conditions represented in the PU dataset.

\begin{figure}[ht!]
\centering
\includegraphics[width=0.95\textwidth]{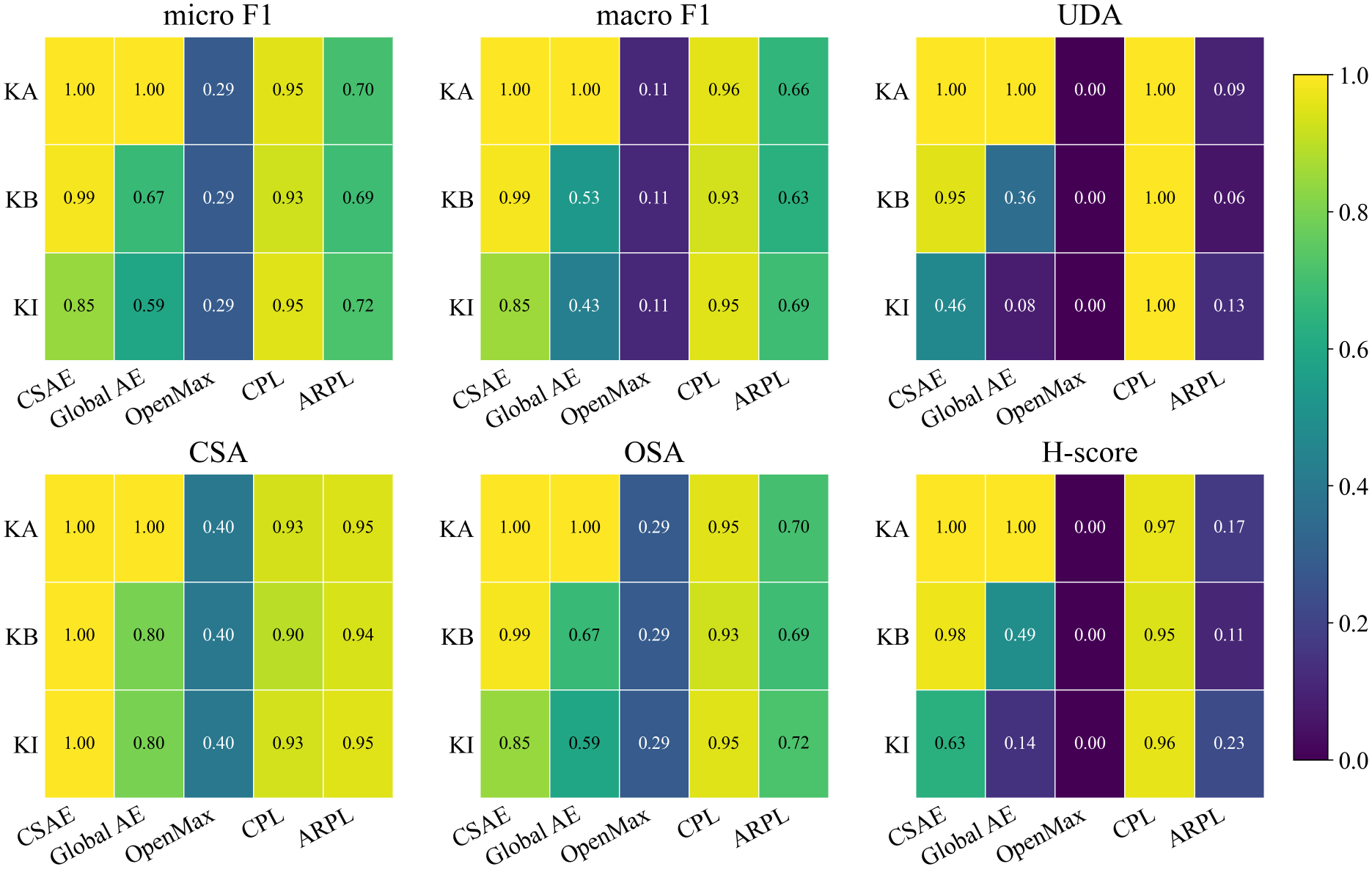}
\caption{Task-wise performance heatmaps for fault type diagnosis on the PU dataset. Rows indicate the unknown fault type, and columns indicate the compared methods.}
\label{fig:score_PU_type}
\end{figure}

\begin{figure}[ht!]
\centering
\includegraphics[width=0.95\textwidth]{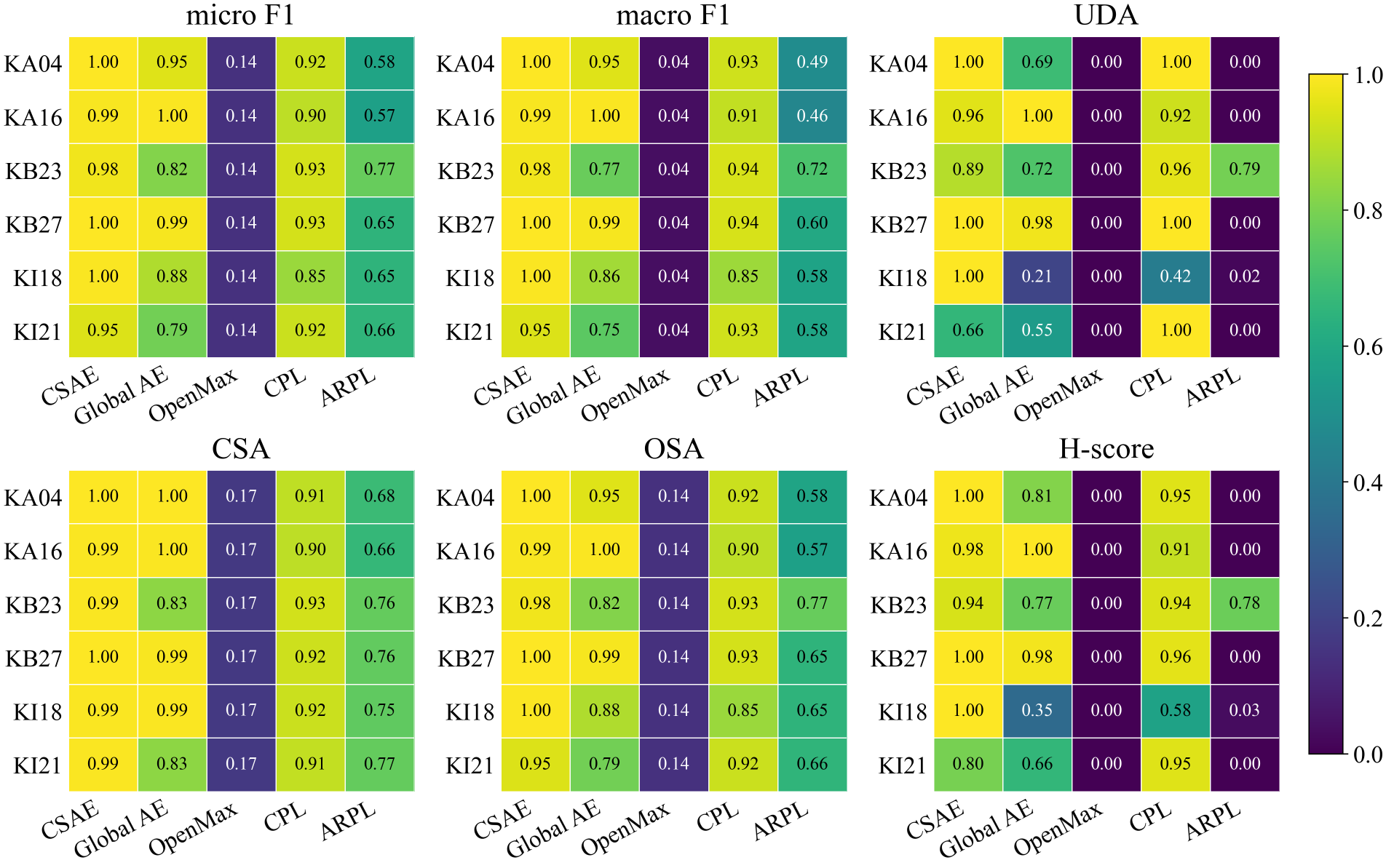}
\caption{Task-wise performance heatmaps for fault severity diagnosis on the PU dataset. Rows indicate the unknown fault severity, and columns indicate the compared methods.}
\label{fig:score_PU_severity}
\end{figure}

Figs.~\ref{fig:score_PU_type} and \ref{fig:score_PU_severity} show the task-specific diagnostic performance on the PU dataset. In fault-type diagnosis, the proposed method maintains high closed-set accuracy (CSA), but its unknown-detection accuracy (UDA) decreases when the inner-race fault type (KI) is treated as unknown, resulting in a lower H-score. CPL achieves the highest average H-score in this setting.

In fault-severity diagnosis, the proposed method delivers stable performance across most unknown severity states and maintains a favorable balance between UDA and CSA, resulting in consistently high OSA and H-scores. Despite a performance decrease for KI21, the proposed method achieves the highest average H-score in Table~\ref{tab:Result_P}. CPL also performs well in several tasks but shows a larger decrease for KI18. These results support the effectiveness of class-specific reconstruction for fine-grained severity diagnosis, where adjacent damage states may form highly overlapping manifolds.


\subsection{Computational efficiency of MGDC-based STFT configuration selection}  \label{sec:Cost}

This section evaluates the computational efficiency of the proposed MGDC strategy for STFT configuration selection. The H-score values reported in this subsection are computed on the selection set. Since STFT configuration strongly affects diagnostic performance, exhaustive search requires training and validating the diagnostic network for every candidate configuration, which becomes costly as the search space grows. We therefore compare MGDC with exhaustive search and representative hyperparameter optimization methods, including TPE, Hyperband, and BOHB.

\begin{table}[ht!]
\centering
\caption{Comparison of STFT configuration selection strategies for cumulative computational time and best H-score. Exhaustive search evaluates all candidate configurations, whereas MGDC evaluates candidates sequentially along the Silhouette-ranked candidate-evaluation path. Specifically, MGDC ($R=r$) denotes the best H-score obtained after validating the top-$r$ STFT configurations. The HPO-based methods include TPE, Hyperband, and BOHB.}
\label{tab:Time_comparison}
\begin{tabular}{clccc}
\toprule
    \textbf{Dataset} & \textbf{Strategy} & \textbf{Computational time (s)} & \textbf{Speedup} & \textbf{Best H-score} \\
\midrule
    \multirow{7}{*}{CWRU} 
    & Exhaustive Search   &  8297.56 & -              & 0.9948 \\
    & MGDC ($R=1$)        &   147.22 & $56.36\times$  & 0.9323 \\
    & MGDC ($R=9$)        &  1457.04 & $5.69\times$   & 0.9948 \\
    & TPE (intermediate)  &   123.86 & $66.99\times$  & 0.9729 \\
    & TPE (final)         &  7604.29 & $1.09\times$   & 0.9948 \\
    & Hyperband           &   941.64 & $8.81\times$   & 0.9323 \\
    & BOHB                &   923.06 & $8.99\times$   & 0.9729 \\
    \midrule
    \multirow{5}{*}{PU} 
    & Exhaustive Search   & 25853.00 & -              & 0.9883 \\
    & MGDC ($R=2$)        &   865.59 & $29.87\times$  & 0.9883 \\
    & TPE                 & 18979.82 & $1.36\times$   & 0.9883 \\
    & Hyperband           &  6091.80 & $4.24\times$   & 0.9790 \\
    & BOHB                &  3375.29 & $7.66\times$   & 0.9871 \\
\bottomrule
\end{tabular}
\end{table}

Table~\ref{tab:Time_comparison} summarizes the computational time and best H-score obtained by each STFT configuration selection strategy. HPO-based methods require repeated network-based feedback during the search process. In contrast, MGDC first ranks the STFT configurations using the training-free Silhouette score and then validates the ranked candidates sequentially. Thus, MGDC acts as an anytime screening strategy: increasing $R$ expands the validation budget and can improve the best-found H-score.
For the CWRU dataset, exhaustive search requires 8297.56 seconds and achieves a best H-score of 0.9948. MGDC obtains an H-score of 0.9323 using only the top-1 configuration in 147.22 seconds and reaches the same best H-score as exhaustive search at $R=9$ within 1457.04 seconds. This substantially reduces the computational cost while preserving the best diagnostic performance. In comparison, TPE requires 7604.29 seconds and achieves an H-score of 0.9948, while Hyperband and BOHB require 941.64 and 923.06 seconds, respectively. Although Hyperband and BOHB reduce the search time, their best H-scores remain lower than that obtained by MGDC.

The PU dataset further illustrates the budget-controllable behavior of MGDC. Exhaustive search requires 25853.00 seconds to obtain the best H-score of 0.9883, and TPE requires 18979.82 seconds to reach the same H-score. In contrast, MGDC reaches the same best H-score of 0.9883 at $R=2$ within 865.59 seconds. This result shows that MGDC can identify the globally best STFT configurations by evaluating high Silhouette-ranked candidates. Compared with TPE, MGDC reaches the same best H-score with lower computational time. Compared with Hyperband and BOHB, MGDC achieves a higher final H-score with lower computational time.

\begin{figure}[ht!]
\centering 
\begin{subfigure}[b]{0.48\textwidth} 
    \includegraphics[width=\textwidth]{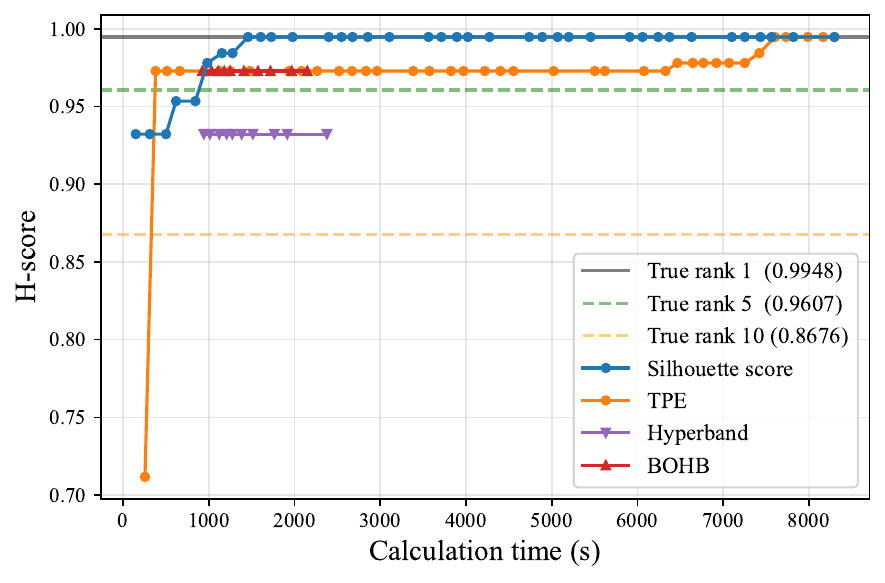}
    \caption{}
    \label{fig:CWRU_opt} 
\end{subfigure}
\hfill
\begin{subfigure}[b]{0.48\textwidth} 
    \includegraphics[width=\textwidth]{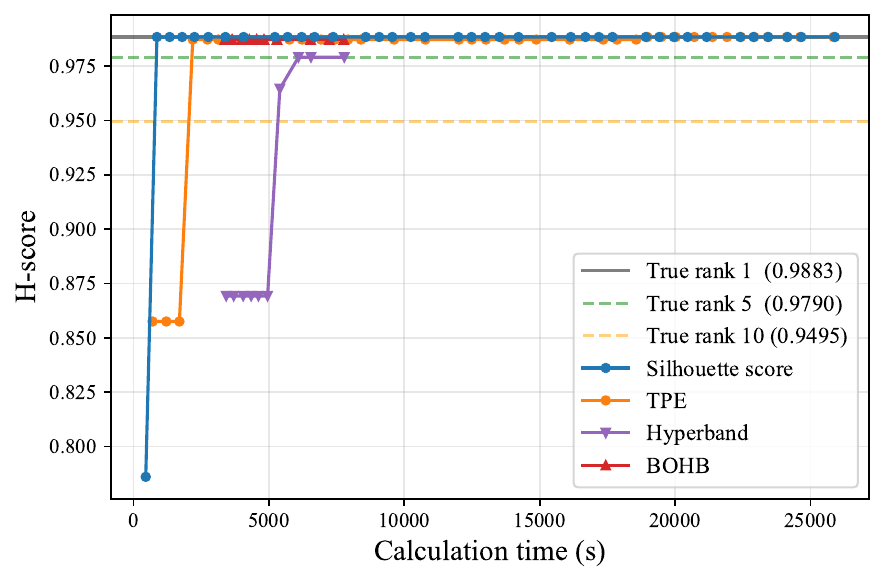}
    \caption{}
    \label{fig:PU_opt} 
\end{subfigure}
\caption{Anytime comparison of STFT configuration selection strategies. The x-axis denotes cumulative calculation time, and the y-axis denotes the best H-score found up to each time point. The horizontal reference lines indicate the H-score levels of the true rank-1, rank-5, and rank-10 configurations obtained from the full search space.}
\label{fig:opt}
\end{figure}

Fig.~\ref{fig:opt} provides an anytime comparison of the search trajectories. The x-axis represents cumulative calculation time, and the y-axis represents the best H-score found up to each time point. On both datasets, MGDC rapidly enters a high-H-score region because the Silhouette ranking prioritizes STFT configurations with strong representation-level separability before network training. For the CWRU dataset, MGDC reaches the global-best H-score much earlier than exhaustive search and TPE. For the PU dataset, MGDC first obtains a competitive H-score with only a small number of validated candidates and then reaches the global-best H-score as the validation budget increases. These results indicate that MGDC is effective as a lightweight front-end screening strategy for STFT configuration selection. The proposed method reduces unnecessary network-training trials while still allowing the search to be refined when additional computational budget is available.


\subsection{Threshold sensitivity analysis} \label{sec:Threshold}

The proposed method rejects unknown samples using two class-specific criteria: a dimension-wise latent boundary and a reconstruction-error threshold. The confidence percentile $\alpha$ controls both criteria and therefore determines the balance between accepting known samples and rejecting unknown samples. We evaluate $\alpha\in\{0.9,0.95,0.99,0.9999\}$ on the selection set and use the H-score, the harmonic mean of CSA and UDA, to quantify this balance.

\begin{figure}[ht!]
\centering 
\begin{subfigure}[b]{0.48\textwidth} 
    \includegraphics[width=\textwidth]{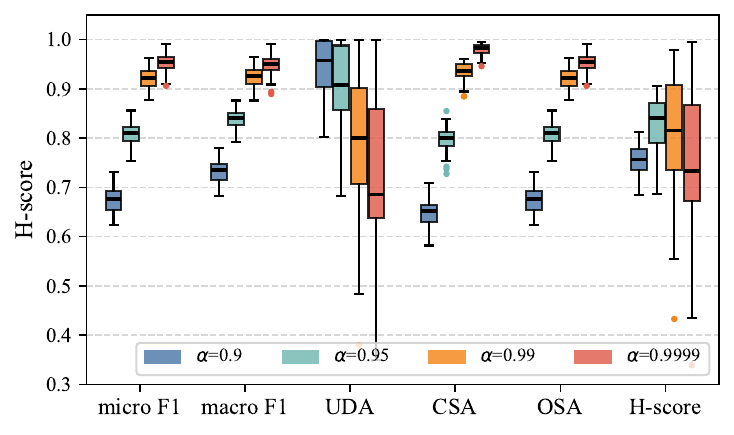}
    \caption{}
    \label{fig:CWRU_th} 
\end{subfigure}
\hfill
\begin{subfigure}[b]{0.48\textwidth} 
    \includegraphics[width=\textwidth]{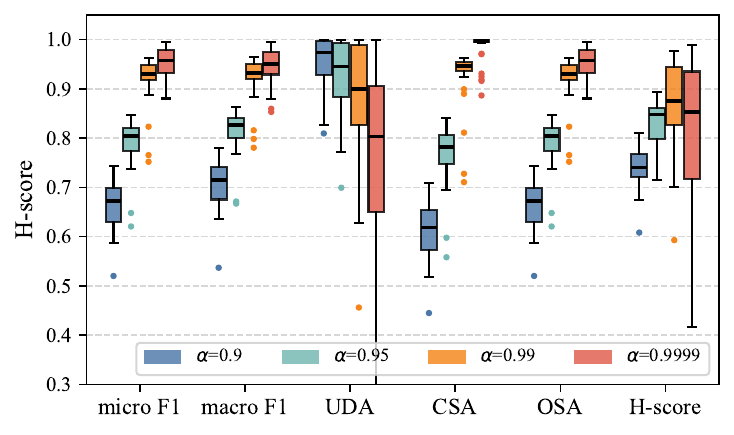}
    \caption{}
    \label{fig:PU_th} 
\end{subfigure}
\caption{Sensitivity of OSFD performance to the confidence percentile $\alpha$ on the CWRU and PU selection sets.}
\label{fig:th}
\end{figure}

Figs.~\ref{fig:CWRU_th} and \ref{fig:PU_th} show the distributions of the diagnostic metrics across all candidate STFT configurations at each value of $\alpha$. A lower percentile, such as $\alpha=0.9$, produces tighter acceptance boundaries and improves UDA but reduces CSA by rejecting more known samples. Increasing $\alpha$ expands the acceptance region and improves CSA, although the resulting relaxation can reduce the consistency of unknown-state detection. The highest H-score is obtained at $\alpha=0.9999$. However, the wide H-score distribution at this value shows that performance under a broad acceptance boundary depends strongly on the quality of the initial time-frequency representation. This result further supports the use of MGDC to select a discriminative STFT configuration before model training.


\subsection{Ablation study} \label{sec:ablation}
We conduct an ablation study on the CWRU and PU test sets to quantify the contribution of each component of the proposed rejection mechanism. The full CSAE model combines dimension-wise latent boundaries and reconstruction-error thresholds, both calibrated separately for each known class. Four variants are considered. LatentOnly uses only the latent-boundary criterion, whereas ReconOnly uses only the reconstruction-error criterion. RejectAND rejects a sample only when both criteria identify the sample as unknown; in contrast, the full model rejects a sample when either acceptance criterion is violated. GlobalTh replaces the class-specific thresholds with thresholds shared across all known classes. These variants evaluate the contributions of dual-criteria rejection, fusion logic, and class-specific threshold calibration.

\begin{table}[ht!]
\centering 
\caption{Ablation results for the proposed rejection mechanism of the CWRU dataset.}
\label{tab:Ablation_C}
\begin{tabular}{c l ccccc}
\toprule
    \textbf{Granularity} & \textbf{Metric} & \textbf{CSAE full} & \textbf{LatentOnly} & \textbf{ReconOnly} & \textbf{RejectAND} & \textbf{GlobalTh} \\
\midrule
    \multirow{6}{*}{\shortstack{Fault \\ type}} 
    & micro F1 & 0.9416 & 0.9077 & 0.8799 & 0.8459 & 0.8751 \\
    & macro F1 & 0.9464 & 0.9138 & 0.8939 & 0.8565 & 0.8808 \\
    & UDA      & 0.8242 & 0.7062 & 0.6049 & 0.4869 & 0.5960 \\
    & CSA      & 0.9920 & 0.9941 & 0.9977 & 0.9998 & 0.9947 \\
    & OSA      & 0.9416 & 0.9077 & 0.8799 & 0.8459 & 0.8751 \\
    & H-score  & 0.8892 & 0.8146 & 0.7474 & 0.6463 & 0.7389 \\
\midrule
    \multirow{6}{*}{\shortstack{Fault \\ severity}} 
    & micro F1 & 0.9902 & 0.9904 & 0.9493 & 0.9495 & 0.9053 \\
    & macro F1 & 0.9904 & 0.9904 & 0.9375 & 0.9371 & 0.8772 \\
    & UDA      & 0.9953 & 0.9681 & 0.5230 & 0.4959 & 0.0644 \\
    & CSA      & 0.9896 & 0.9928 & 0.9967 & 0.9999 & 0.9988 \\
    & OSA      & 0.9902 & 0.9904 & 0.9493 & 0.9495 & 0.9053 \\
    & H-score  & 0.9924 & 0.9786 & 0.5670 & 0.5522 & 0.1080 \\
\bottomrule
\end{tabular}
\end{table}

Table~\ref{tab:Ablation_C} shows that the full CSAE model achieves the highest H-score for both fault type and fault severity diagnosis on the CWRU dataset. In fault type diagnosis, removing either rejection criterion reduces UDA and H-score despite preserving high CSA. The combined criteria therefore improve unknown-fault detection without substantially reducing known-class accuracy.

The differences are more pronounced in fault-severity diagnosis. LatentOnly remains competitive but yields a lower H-score than the full model. ReconOnly and RejectAND maintain high CSA but show substantial reductions in UDA, indicating that reconstruction error alone does not adequately reject unknown severity states. The stronger performance of LatentOnly relative to ReconOnly also shows that the latent-boundary criterion contributes more strongly to unknown-severity detection. GlobalTh produces the largest decrease in UDA and H-score, demonstrating the importance of accounting for class-dependent latent and reconstruction-error distributions.

\begin{table}[ht!]
\centering 
\caption{Ablation results for the proposed rejection mechanism of the PU dataset.}
\label{tab:Ablation_P}
\begin{tabular}{c l ccccc}
\toprule
    \textbf{Granularity} & \textbf{Metric} & \textbf{CSAE full} & \textbf{LatentOnly} & \textbf{ReconOnly} & \textbf{RejectAND} & \textbf{GlobalTh} \\
\midrule
    \multirow{6}{*}{\shortstack{Fault \\ type}} 
    & micro F1 & 0.9435 & 0.9436 & 0.7774 & 0.7775 & 0.7646 \\
    & macro F1 & 0.9470 & 0.9471 & 0.7655 & 0.7655 & 0.7367 \\
    & UDA      & 0.8065 & 0.8060 & 0.2217 & 0.2212 & 0.1785 \\
    & CSA      & 0.9982 & 0.9986 & 0.9996 & 1.0000 & 0.9991 \\
    & OSA      & 0.9435 & 0.9436 & 0.7774 & 0.7775 & 0.7646 \\
    & H-score  & 0.8698 & 0.8695 & 0.3442 & 0.3436 & 0.2885 \\
\midrule
    \multirow{6}{*}{\shortstack{Fault \\ severity}} 
    & micro F1 & 0.9846 & 0.9841 & 0.9422 & 0.9416 & 0.8926 \\
    & macro F1 & 0.9844 & 0.9838 & 0.9322 & 0.9308 & 0.8623 \\
    & UDA      & 0.9182 & 0.9116 & 0.5987 & 0.5921 & 0.2566 \\
    & CSA      & 0.9957 & 0.9961 & 0.9994 & 0.9999 & 0.9987 \\
    & OSA      & 0.9846 & 0.9841 & 0.9422 & 0.9416 & 0.8926 \\
    & H-score  & 0.9509 & 0.9469 & 0.6864 & 0.6768 & 0.3057 \\
\bottomrule
\end{tabular}
\end{table}

Table~\ref{tab:Ablation_P} shows a similar pattern on the PU dataset. In fault-type diagnosis, the full CSAE and LatentOnly variants achieve comparable results, although the full model provides the highest H-score. ReconOnly, RejectAND, and GlobalTh retain high CSA but show large decreases in UDA, indicating poor rejection of unknown fault types.

In fault-severity diagnosis, the full CSAE model achieves the highest H-score. LatentOnly provides the closest performance but shows a moderate reduction in UDA. ReconOnly and RejectAND again maintain high CSA while producing substantially lower UDA, and GlobalTh exhibits the largest overall degradation. The consistent results across both datasets show that latent-boundary checking provides the primary rejection capability, while the reconstruction-error criterion offers complementary information. Class-specific calibration is also necessary to account for differences among known-class distributions.


\section{Conclusion} \label{sec:Conclusion}

We proposed a hybrid OSFD method tailored to fine-grained fault severity diagnosis. To address the preprocessing bottleneck, MGDC screened candidate STFT configurations using the Silhouette score and validated only a reduced candidate set. This procedure identified effective spectrogram representations while reducing the cost of exhaustive network-based search. Using the selected representations, the proposed architecture combined a shared feature extractor with a bank of CSAEs. Each CSAE corresponds to one known severity class and constructs a compact class-specific latent manifold. The dual-criteria rejection mechanism combines latent dimension-wise boundaries with class-specific reconstruction error, enabling simultaneous known-class classification and unknown fault rejection.

Experiments on the CWRU and PU datasets demonstrated the effectiveness and computational efficiency of the proposed method. The method did not consistently outperform the strongest baselines in coarse-grained fault type diagnosis, where fault classes are generally more separable. In contrast, the proposed method showed its main advantage in fine-grained fault severity diagnosis, where adjacent degradation states form densely entangled feature manifolds. In this setting, the method achieved balanced performance for both known-class classification and unknown fault rejection. MGDC also reduced the computational cost of STFT configuration selection by limiting network-based validation to a small set of high-scoring candidates.

This study has three main limitations. The experiments focus on bearing datasets, MGDC uses known labels during STFT screening, and the CSAE structure is more effective for fine-grained severity diagnosis than for all OSFD settings. Future work will evaluate the framework on other rotating machinery components, extend MGDC to weakly labeled or unlabeled data, and investigate incremental and domain-adaptive CSAE models for online monitoring under variable operating conditions.

\section*{Acknowledgment} 
This work was supported by the Institute of Information \& Communications Technology Planning \& Evaluation (IITP) grant funded by the Korea government (MSIT) (No. RS-2025-02304285, Development of Digital Triplet Based Predictive Maintenance Solution for Future Power Facility Innovation).

\bibliography{citation}
\bibliographystyle{unsrt}


\renewcommand\theequation{\Alph{section}\arabic{equation}} 
\counterwithin*{equation}{section} 
\renewcommand\thefigure{\Alph{section}\arabic{figure}} 
\counterwithin*{figure}{section} 
\renewcommand\thetable{\Alph{section}\arabic{table}} 
\counterwithin*{table}{section} 

\begin{appendices}

\section{Detailed OSFD evaluation metrics} \label{apdx:Detailed OSFD evaluation metrics}

This section details the evaluation metrics: micro F1-score, macro F1-score, unknown detection accuracy (UDA), closed-set accuracy (CSA), open-set accuracy (OSA), and the harmonic score (H-score).

To provide a balanced assessment of diagnostic precision and sensitivity, we utilize the F1-score, which serves as the harmonic mean of precision ($\mathrm{Pr}$) and recall ($\mathrm{Re}$). We define the fundamental components---true positives ($\mathrm{TP}$), true negatives ($\mathrm{TN}$), false positives ($\mathrm{FP}$), and false negatives ($\mathrm{FN}$)---to derive these measures as follows:
\begin{align}
    \mathrm{Pr} = \frac{\mathrm{TP}}{\mathrm{TP}+\mathrm{FP}},  \quad
    \mathrm{Re} = \frac{\mathrm{TP}}{\mathrm{TP}+\mathrm{FN}},  \quad
    \mathrm{F1} = 2\times \frac{\mathrm{Pr} \times \mathrm{Re}}{\mathrm{Pr} +\mathrm{Re}}.
\end{align}

For a comprehensive evaluation across multiple fault categories, we implement both micro and macro variants of the F1-score. The micro F1-score ($\mathrm{F1}^{\mu}$) calculates the global average by aggregating the $\mathrm{TP}_i$, $\mathrm{TN}_i$, $\mathrm{FP}_i$, and $\mathrm{FN}_i$ across all classes, thereby serving as an indicator of overall system effectiveness:
\begin{align}
    \mathrm{Pr}^{\mu} = \frac{\sum_{k=1}^{K+U} \mathrm{TP}_k}{\sum_{k=1}^{K+U} (\mathrm{TP}_k + \mathrm{FP}_k)}, \quad 
    \mathrm{Re}^{\mu} = \frac{\sum_{k=1}^{K+U} \mathrm{TP}_k}{\sum_{k=1}^{K+U} (\mathrm{TP}_k + \mathrm{FN}_k)}, \quad 
    \mathrm{F1}^{\mu} = 2 \times \frac{\mathrm{Pr}^{\mu} \times \mathrm{Re}^{\mu}}{\mathrm{Pr}^{\mu} + \mathrm{Re}^{\mu}},
\end{align}
where $K$ and $U$ are the number of known and unknown classes, respectively.

Conversely, the macro F1-score ($\mathrm{F1}^{M}$) evaluates the model's reliability for each class independently before deriving an unweighted mean. By assigning equal importance to each fault category regardless of sample frequency, the macro F1-score ensures that the diagnostic performance for rare or critical fault severities is accurately captured:
\begin{align}
    \mathrm{F1}^{M} = \frac{1}{K+U} \sum_{k=1}^{K+U} \mathrm{F1}_{k},
\end{align}
where $\mathrm{F1}_{k}$ denotes the F1-score of the $k$th class.

To specifically evaluate the OSFD capability, we employ four specialized metrics: UDA, CSA, OSA, H-score. UDA evaluates the model's success in correctly identifying the designated unknown fault samples. CSA measures the classification accuracy exclusively for known classes, indicating the model's reliability in seen operational states. OSA provides a global accuracy metric by encompassing both known and unknown classes across the entire evaluation set. H-score, representing the harmonic mean of CSA and UDA, serves as a primary indicator of balanced OSFD performance by effectively addressing the inherent trade-off between classifying known samples and detecting unknown samples. The metrics are denoted as
\begin{align}
    \mathrm{UDA} &= \frac{\sum_{u=1}^{U} \mathrm{TP}_u}{\sum_{u=1}^{U} (\mathrm{TP}_u + \mathrm{FN}_u)}, \quad
    \mathrm{CSA} = \frac{\sum_{k=1}^{K} \mathrm{TP}_k}{\sum_{k=1}^{K} (\mathrm{TP}_k + \mathrm{FN}_k)}, \\
    \mathrm{OSA} &= \frac{\sum_{k=1}^{K} \mathrm{TP}_k + \sum_{u=1}^{U} \mathrm{TP}_u}{\sum_{k=1}^{K} (\mathrm{TP}_k + \mathrm{FN}_k) + \sum_{u=1}^{U}(\mathrm{TP}_u + \mathrm{FN}_u)}, \quad     \mathrm{H\text{-}score} = 2\times \frac{\mathrm{CSA} \times \mathrm{UDA}}{\mathrm{CSA} + \mathrm{UDA}}, \nonumber \\
\end{align}
where $U$ and $K$ are the number of unknown and known classes, respectively.


\section{Detailed separability metrics for STFT configuration screening} \label{apdx:Metrics}

In addition to the Silhouette score introduced in Section~\ref{sec:Silhouette}, several auxiliary separability metrics were evaluated to assess class-wise compactness and inter-class separability. Let $\mathbf{x}_i \in \mathbb{R}^{d}$ denote the $i$th feature vector, where $i=1,\dots,N$, $d$ is the feature dimension, and $N$ is the number of samples. Let $C_k$ be the set of samples belonging to the $k$th known class, where $k=1,\dots,K$ and $K$ is the number of known classes. Here, $|C_k|$ denotes the number of samples in set $C_k$. $k_1$ and $k_2$ denote two different class indices used for pairwise class comparisons. The centroid of class $k$ is denoted by $\boldsymbol{\mu}_k$, and the global centroid is denoted by $\boldsymbol{\mu}$. A small constant $\epsilon$ is used for numerical stability. These metrics were used only for preprocessing-level analysis before network training. For metrics where smaller values indicate better separability, such as the Davies-Bouldin Index, the ranking direction was reversed when comparing candidate representations.

Mean entropy measures the average uncertainty or energy dispersion of normalized spectrogram representations based on Shannon entropy~\citep{shannon1948mathematical}:
\begin{align}
    H(\mathbf{x}_i) &= -\sum_{r=1}^{d} p_{i,r}\log(p_{i,r} +\epsilon), \\
    p_{i,r} &= \frac{x_{i,r}}{\sum_{\xi=1}^{d}x_{i,\xi} +\epsilon},
\end{align}
where $p_{i,r}$ denotes the normalized contribution of the $r$th feature component of $\mathbf{x}_i$ and $d$ is the feature dimension. Lower entropy generally indicates a more concentrated representation. The final mean entropy score was obtained by averaging $H(\mathbf{x}_i)$ over all samples.

Calinski-Harabasz Index evaluates the ratio between-class scatter and within-class scatter with degree-of-freedom normalization~\citep{calinski1974dendrite}:
\begin{align}
    \mathrm{CH} = \frac{\mathrm{tr}(\mathbf{S}_B)/(K-1)}{\mathrm{tr}(\mathbf{S}_W)/(N-K)},
\end{align}
where $N$ is the number of samples, $K$ is the number of classes, $\mathrm{tr}(\cdot)$ denotes the trace operator, and $\mathbf{S}_B$ and $\mathbf{S}_W$ are the between-class and within-class scatter matrices, respectively:
\begin{align}
    \mathbf{S}_B &= \sum_{k=1}^{K} |C_k|(\boldsymbol{\mu}_k-\boldsymbol{\mu})(\boldsymbol{\mu}_k-\boldsymbol{\mu})^{\top}, \\
    \mathbf{S}_W &= \sum_{k=1}^{K} \sum_{\mathbf{x}_i \in C_k}(\mathbf{x}_i-\boldsymbol{\mu}_k)(\mathbf{x}_i-\boldsymbol{\mu}_k)^{\top}.
\end{align}
A larger value indicates better class separability. 

The variance ratio uses the same principle but directly compares between-class and within-class variance without the same normalization. The variance ratio was computed as
\begin{align}
    \mathrm{VR} = \frac{\mathrm{tr}(\mathbf{S}_B)}{\mathrm{tr}(\mathbf{S}_W) +\epsilon}.
\end{align}

The centroid cosine distance evaluates angular separation between centroids of classes $k_1$ and $k_2$:
\begin{align}
    \mathrm{CCD} = \frac{2}{K(K-1)}\sum_{k_1<k_2} \left(1 - \frac{\boldsymbol{\mu}_{k_1}^{\top}\boldsymbol{\mu}_{k_2}} {\|\boldsymbol{\mu}_{k_1}\|_2\|\boldsymbol{\mu}_{k_2}\|_2 +\epsilon}\right).
\end{align}
A larger value indicates that the class centroids are more directionally separated.

The Bhattacharyya distance measures the overlap between two class distributions~\citep{bhattacharyya1943measure}. Assuming Gaussian class distributions, the pairwise distance between classes $k_1$ and $k_2$ is computed as
\begin{align}
    D_B(k_1,k_2) = \frac{1}{8} (\boldsymbol{\mu}_{k_1}-\boldsymbol{\mu}_{k_2})^{\top} \boldsymbol{\Sigma}^{-1} (\boldsymbol{\mu}_{k_1}-\boldsymbol{\mu}_{k_2}) + \frac{1}{2} \ln \frac{|\boldsymbol{\Sigma}|}{\sqrt{|\boldsymbol{\Sigma}_{k_1}||\boldsymbol{\Sigma}_{k_2}|}},
\end{align}
where $\boldsymbol{\Sigma}_{k_1}$ and $\boldsymbol{\Sigma}_{k_2}$ denote the covariance matrices of classes $k_1$ and $k_2$, respectively, and $\boldsymbol{\Sigma}=(\boldsymbol{\Sigma}_{k_1}+\boldsymbol{\Sigma}_{k_2})/2$. The final Bhattacharyya distance score was obtained by averaging $D_B(k_1,k_2)$ over all class pairs. In practice, a small diagonal regularization term was added to the covariance matrices for numerical stability. A larger distance indicates smaller distributional overlap. 

The Dunn Index measures the ratio between the minimum inter-class distance and the maximum intra-class diameter~\citep{dunn1974well}. A larger value indicates compact within-class distributions and well-separated class clusters:
\begin{align}
    \mathrm{DI} = \frac{\min_{k_1 \neq k_2} \delta(C_{k_1},C_{k_2})}{\max_k \Delta(C_k)},
\end{align}
where $\delta(C_{k_1},C_{k_2})$ denotes the minimum pairwise distance between samples from two different classes, and $\Delta(C_k)$ denotes the maximum pairwise distance within class $C_k$.

The inter/intra ratio directly compares the average distance between class centroids with the average within-class dispersion:
\begin{align}
    \mathrm{IIR} = 
    \frac{\frac{2}{K(K-1)}\sum_{k_1<k_2}\|\boldsymbol{\mu}_{k_1}-\boldsymbol{\mu}_{k_2}\|_2}
    {\frac{1}{K}\sum_{k=1}^{K}\frac{1}{|C_k|}\sum_{\mathbf{x}_i \in C_k}\|\mathbf{x}_i-\boldsymbol{\mu}_k\|_2}.
\end{align}
A larger value indicates better class separability.

The Fisher score measures class separability by comparing between-class variance with within-class variance~\citep{fisher1936use}. It is computed feature-wise and averaged over all dimensions:
\begin{align}
    \mathrm{FS} = \frac{1}{d} \sum_{r=1}^{d} \frac{\sum_{k=1}^{K}|C_k|(\mu_{k,r}-\mu_r)^2} {\sum_{k=1}^{K}|C_k|\sigma_{k,r}^2 +\epsilon},
\end{align}
where $\mu_{k,r}$ and $\sigma_{k,r}^{2}$ denote the mean and variance of the $r$th feature dimension in class $k$, respectively, and $\mu_r$ denotes the global mean of the $r$th feature dimension. A larger value indicates that the feature dimension provides stronger class discrimination.

The Davies-Bouldin Index measures the average worst-case similarity between each class and its most similar neighboring class~\citep{davies1979cluster}. It is defined as
\begin{align}
    \mathrm{DBI} =
    \frac{1}{K}\sum_{k=1}^{K}
    \max_{k_2\neq k}
    \frac{s_k+s_{k_2}}{\|\boldsymbol{\mu}_k-\boldsymbol{\mu}_{k_2}\|_2 +\epsilon},
\end{align}
where $s_k=\frac{1}{|C_k|}\sum_{\mathbf{x}_i\in C_k}\|\mathbf{x}_i-\boldsymbol{\mu}_k\|_2$ denotes the average within-class dispersion of class $k$. A smaller value indicates better separation.


\section{Detailed metric-guided data-centric configuration search strategy for preprocessing} \label{apdx:MGDC}

\begin{figure}[ht!]
\centering 
\includegraphics[width=0.99\linewidth]{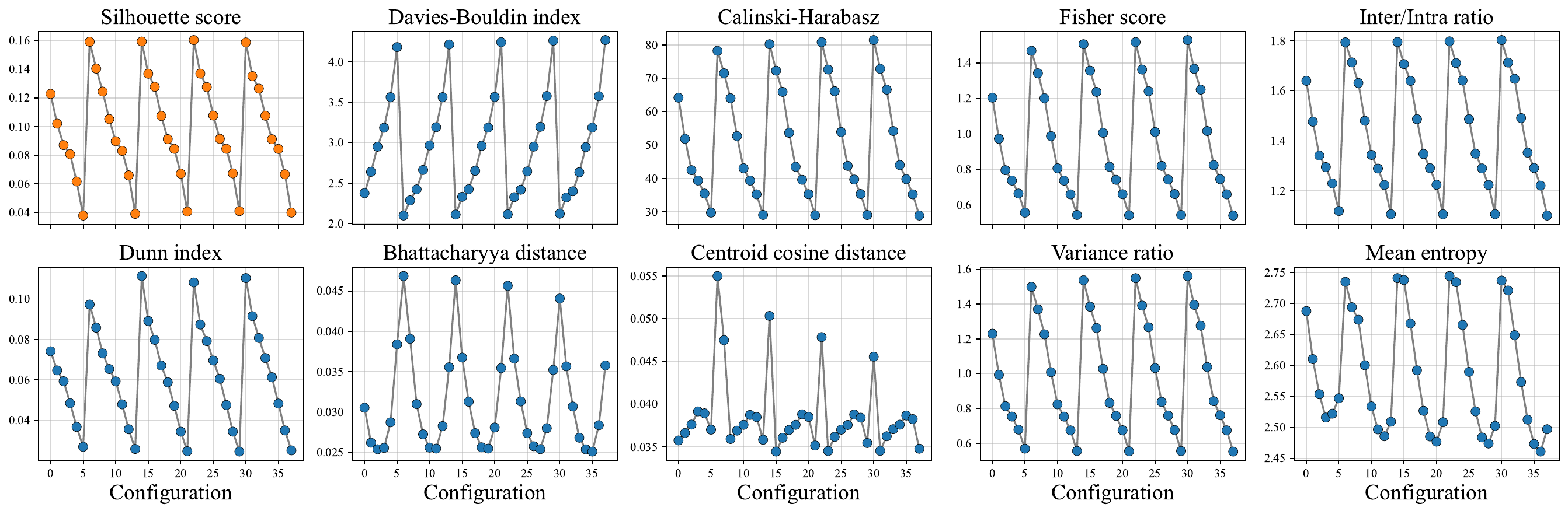}
\caption{Distribution of the clustering metrics across all STFT configurations on the CWRU dataset} 
\label{fig:CWRU_metrics} 
\end{figure}

\begin{figure}[ht!]
\centering 
\includegraphics[width=0.99\linewidth]{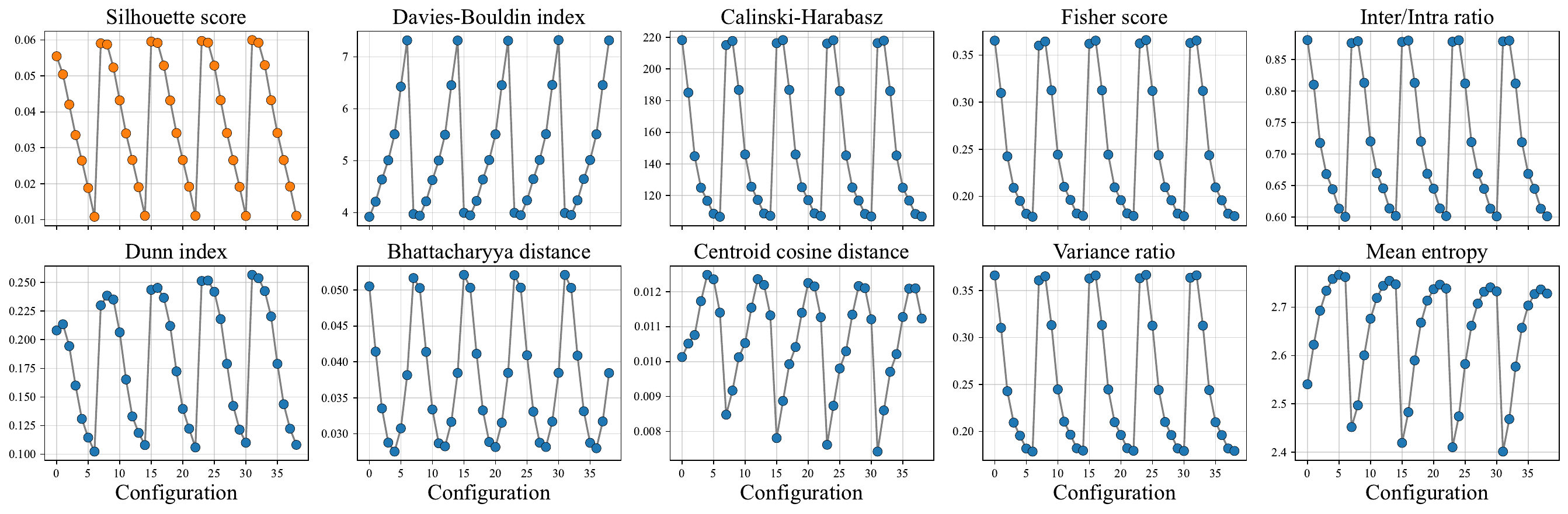}
\caption{Distribution of the clustering metrics across all STFT configurations on the PU dataset} 
\label{fig:PU_metrics} 
\end{figure}

\begin{figure}[ht!]
\centering 
\includegraphics[width=0.8\linewidth]{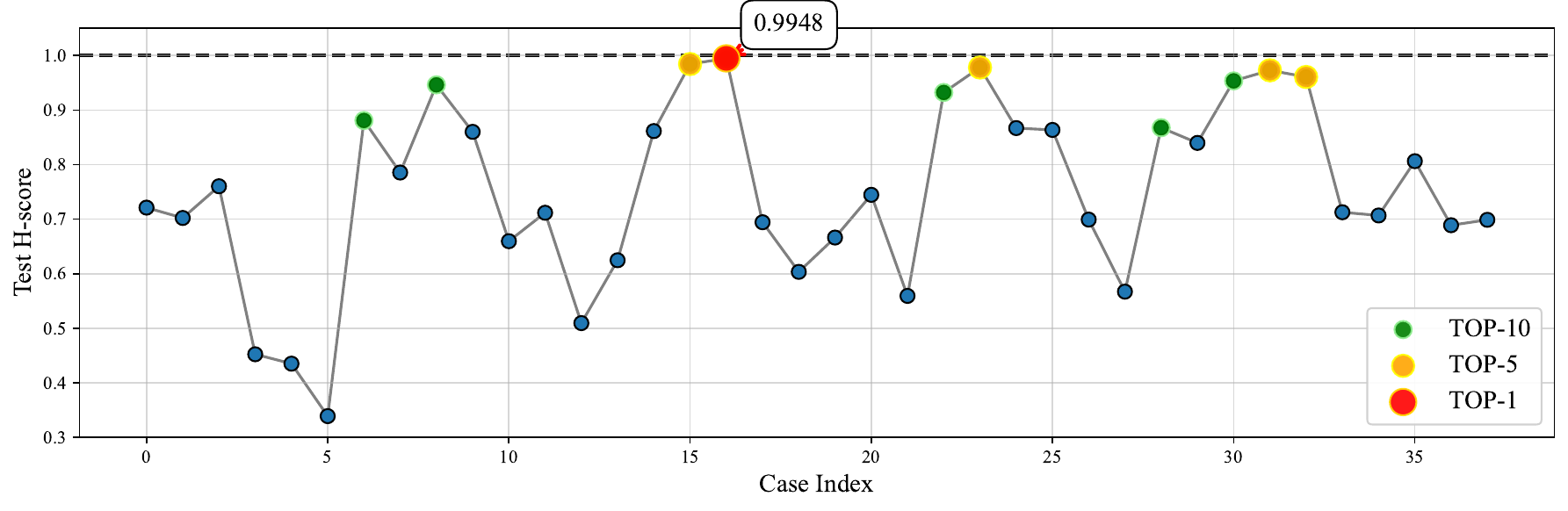}
\caption{H-score trends on the CWRU dataset} 
\label{fig:CWRU_acc} 
\end{figure}

\begin{figure}[ht!]
\centering 
\includegraphics[width=0.8\linewidth]{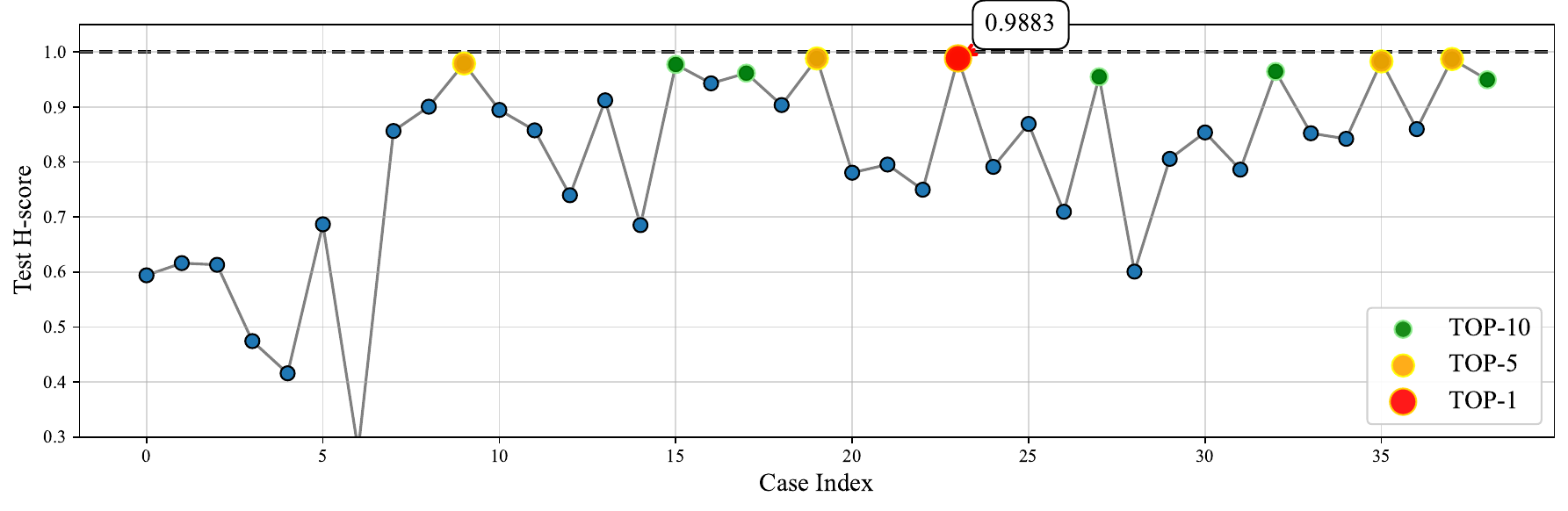}
\caption{H-score corresponding to each STFT configuration on the PU dataset} 
\label{fig:PU_acc} 
\end{figure}

\begin{figure}[ht!]
\centering 
\includegraphics[width=0.8\linewidth]{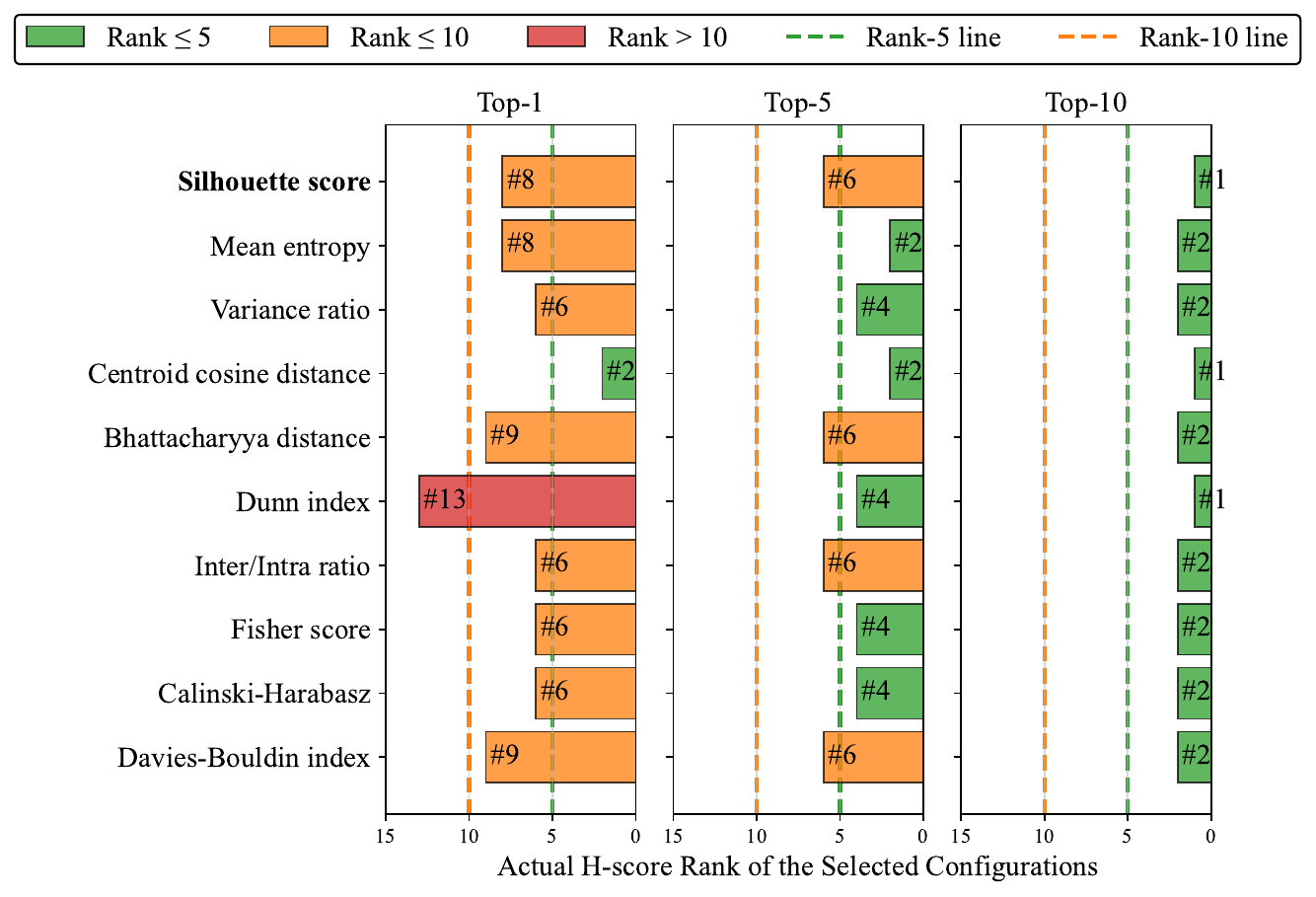}
\caption{Rank evaluation of metric-based STFT configuration screening on the CWRU dataset. The results show that top-1 selection can vary depending on the metric, whereas a reduced candidate set $\mathcal{H}_R$ with a small selection size provides a more reliable way to include diagnostically effective STFT configurations.} 
\label{fig:CWRU_avg_rank} 
\end{figure}

\begin{figure}[ht!]
\centering 
\includegraphics[width=0.8\linewidth]{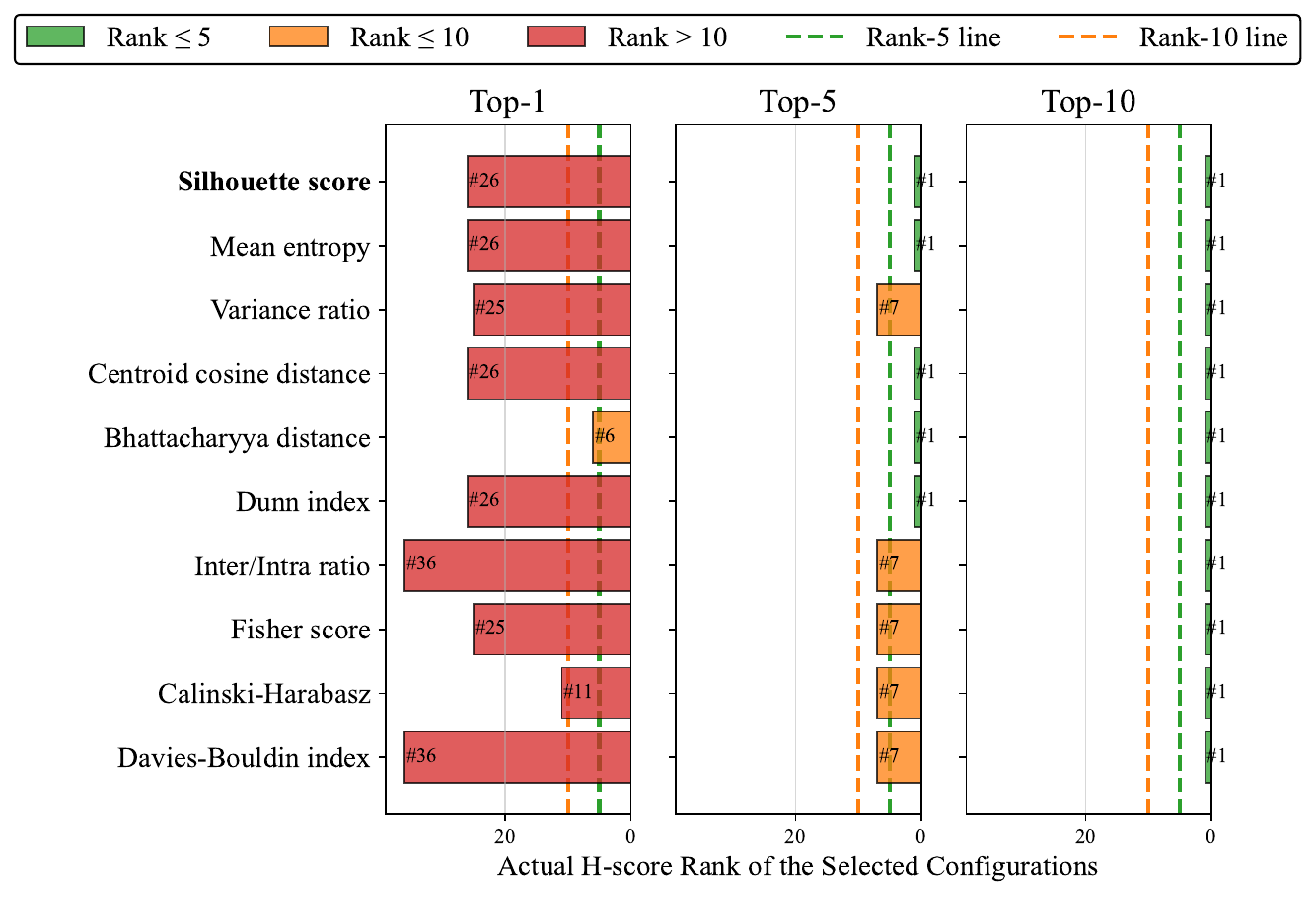} 
\caption{Rank evaluation of metric-based STFT configuration screening on the PU dataset. The results confirm that selecting a small reduced candidate set $\mathcal{H}_R$, rather than relying only on the top-1 configuration, improves the robustness of STFT configuration selection.} 
\label{fig:PU_avg_rank} 
\end{figure}

Fig.s~\ref{fig:CWRU_metrics} through \ref{fig:PU_avg_rank} validate the metric-guided data-centric (MGDC) candidate screening strategy. Specifically, Fig.s~\ref{fig:CWRU_metrics} and \ref{fig:PU_metrics} illustrate the distribution of the proposed objective clustering metrics across all evaluated STFT configurations. To verify whether these metrics genuinely reflect the final diagnostic performance, the actual downstream classification accuracies (H-scores) for each configuration are plotted in Fig.s~\ref{fig:CWRU_acc} and \ref{fig:PU_acc}. 

Fig.s~\ref{fig:CWRU_avg_rank} and \ref{fig:PU_avg_rank} provide additional evidence for the reliability of metric-guided Short-Time Fourier Transform (STFT) configuration screening. The results show that relying only on the single highest-ranked candidate STFT configuration, i.e., $R=1$, can be unstable depending on the selected separability metric. Although some metrics successfully assign high ranks to diagnostically effective configurations, other metrics select suboptimal configurations, resulting in noticeable degradation in the H-score. Therefore, the final STFT configuration should not be determined solely by the single maximum value of a metric.

To address this instability, the proposed MGDC strategy uses the metric score as a screening criterion rather than as a direct optimization objective. Specifically, candidate STFT configurations $h_j \in \mathcal{H}$ are first ranked according to the separability of their corresponding normalized spectrogram representations. Then, only the top-$R$ ranked configurations are retained as the reduced candidate set $\mathcal{H}_R$, and the selected STFT configuration $h^*$ is determined by H-score within $\mathcal{H}_R$. As shown in the main experiments, a small reduced candidate set is sufficient to include near-optimal STFT configurations for both the CWRU and PU datasets. These results support the main conclusion that the MGDC strategy substantially reduces the number of network-training trials while preserving near-optimal diagnostic performance.    

\end{appendices}


\end{document}

%% file: ListPackages.tex
\usepackage[margin = 1.2in]{geometry}
\usepackage[T1]{fontenc}
\usepackage[english]{babel}
\usepackage[latin1]{inputenc}
\usepackage{mathtools}
\usepackage{amsmath}
\usepackage{amsfonts}
\usepackage{amsthm}
\usepackage{amssymb}
\usepackage{array}
\usepackage{subcaption}
\usepackage{siunitx}

\usepackage{color}
\usepackage{url}
\usepackage{cleveref}
\usepackage{graphicx}
\usepackage{breakcites}
\usepackage{soul} 
\usepackage{enumitem}
\usepackage[title,titletoc,toc]{appendix}



%% file: mymacros.tex
\newtheorem{problem}{Problem}

{\noindent {\textbf{Proof}:} }%
{\hfill $\Box$ \\[1ex] }

\newcommand{\bit}{\begin{itemize}}
\newcommand{\eit}{\end{itemize}}
\newcommand{\ben}{\begin{enumerate}}
\newcommand{\een}{\end{enumerate}}

